\UseRawInputEncoding
\documentclass[
superscriptaddress,
 amsmath,amssymb,
 aps,
 pra,
 twocolumn
]{revtex4-2}

\usepackage[utf8]{inputenc}
\usepackage[T1]{fontenc}
\usepackage{graphicx}
\usepackage{array}
\usepackage{braket}
\usepackage{svg}
\usepackage{csquotes}
\usepackage{xspace}

\usepackage{booktabs}
\usepackage{multirow}
\usepackage{color,soul}
\usepackage{mathtools}
\usepackage[hypertexnames=false]{hyperref} 
\hypersetup{
	colorlinks,
	linkcolor={blue},
	citecolor={blue},
	urlcolor={blue}
}

\usepackage[capitalise]{cleveref}

\usepackage{siunitx}

\newcommand{\ankaaTM}{Ankaa\texttrademark-2\xspace}
\newcommand{\ankaa}{Ankaa-2\xspace}

\begin{document}

\title{Demonstrating real-time and low-latency quantum error correction with superconducting qubits}
\author{Laura Caune}
\email{laura.caune@riverlane.com}
\affiliation{Riverlane, Cambridge, CB2 3BZ, UK}
\author{Luka Skoric}
\email{luka.skoric@riverlane.com}
\affiliation{Riverlane, Cambridge, CB2 3BZ, UK}
\author{Nick S. Blunt}
\email{nick.blunt@riverlane.com}
\author{Archibald Ruban}
\affiliation{Riverlane, Cambridge, CB2 3BZ, UK}
\author{Jimmy McDaniel}
\affiliation{Rigetti Computing, 775 Heinz Avenue, Berkeley, California 94710, USA}
\author{Joseph A. Valery}
\affiliation{Rigetti Computing, 775 Heinz Avenue, Berkeley, California 94710, USA}
\author{Andrew D. Patterson}
\affiliation{Rigetti UK Ltd, 138 Holborn, London, EC1N 2SW, UK}
\author{Alexander V. Gramolin}
\affiliation{Riverlane, Cambridge, Massachusetts 02142, USA}
\author{Joonas Majaniemi}
\affiliation{Riverlane, Cambridge, CB2 3BZ, UK}
\author{Kenton M. Barnes}
\affiliation{Riverlane, Cambridge, CB2 3BZ, UK}
\author{Tomasz Bialas}
\affiliation{Riverlane, Cambridge, CB2 3BZ, UK}
\author{Okan Bu\u{g}daycı}
\affiliation{Riverlane, Cambridge, CB2 3BZ, UK}
\author{Ophelia Crawford}
\affiliation{Riverlane, Cambridge, CB2 3BZ, UK}
\author{Gy\"{o}rgy P. Geh\'{e}r}
\affiliation{Riverlane, Cambridge, CB2 3BZ, UK}
\author{Hari Krovi}
\affiliation{Riverlane, Cambridge, Massachusetts 02142, USA}
\author{Elisha Matekole}
\affiliation{Riverlane, Cambridge, Massachusetts 02142, USA}
\author{Canberk Topal}
\affiliation{Riverlane, Cambridge, CB2 3BZ, UK}
\author{Stefano Poletto}
\affiliation{Rigetti Computing, 775 Heinz Avenue, Berkeley, California 94710, USA}
\author{Michael Bryant}
\affiliation{Rigetti Computing, 775 Heinz Avenue, Berkeley, California 94710, USA}
\author{Kalan Snyder}
\affiliation{Rigetti Computing, 775 Heinz Avenue, Berkeley, California 94710, USA}
\author{Neil I. Gillespie}
\affiliation{Riverlane, Cambridge, CB2 3BZ, UK}
\author{Glenn Jones}
\affiliation{Rigetti Computing, 775 Heinz Avenue, Berkeley, California 94710, USA}
\author{Kauser Johar}
\affiliation{Riverlane, Cambridge, CB2 3BZ, UK}
\author{Earl T. Campbell}
\affiliation{Riverlane, Cambridge, CB2 3BZ, UK}
\affiliation{Department of Physics and Astronomy, University of Sheffield, UK}
\author{Alexander D. Hill}
\affiliation{Rigetti Computing, 775 Heinz Avenue, Berkeley, California 94710, USA}

\date{\today}

\thispagestyle{plain}
\pagestyle{plain}
\begin{abstract}
    Quantum error correction (QEC) will be essential to achieve the accuracy needed for quantum computers to realise their full potential. The field has seen promising progress with demonstrations of early QEC and real-time decoded experiments. As quantum computers advance towards demonstrating a universal fault-tolerant logical gate set, implementing scalable and low-latency real-time decoding will be crucial to prevent the backlog problem, avoiding an exponential slowdown and maintaining a fast logical clock rate. Here, we demonstrate low-latency feedback with a scalable FPGA decoder integrated into the control system of a superconducting quantum processor. We perform an 8-qubit stability experiment with up to $25$ decoding rounds and a mean decoding time per round below $1$~\unit{\micro\second}, showing that we avoid the backlog problem even on superconducting hardware with the strictest speed requirements. We observe logical error suppression as the number of decoding rounds is increased. We also implement and time a fast-feedback experiment demonstrating a decoding response time of $9.6$~\unit{\micro\second} for a total of $9$ measurement rounds. The decoder throughput and latency developed in this work, combined with continued device improvements, unlock the next generation of experiments that go beyond purely keeping logical qubits alive and into demonstrating building blocks of fault-tolerant computation, such as lattice surgery and magic state teleportation.
\end{abstract}
\maketitle


\section{Introduction}\label{sec:intro}

\begin{figure*}[t]
    \centering
    \includegraphics[width=\linewidth]{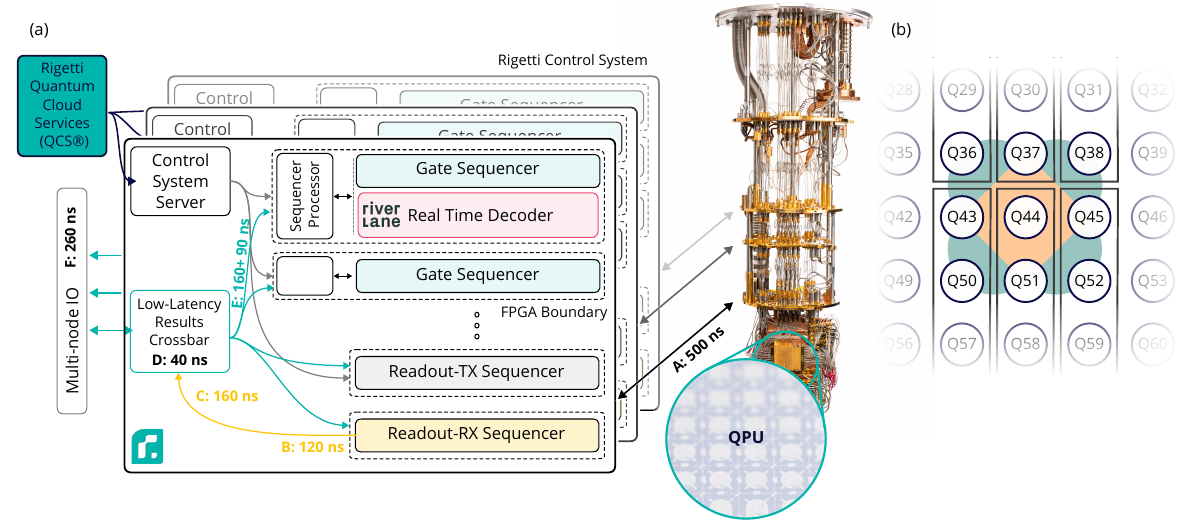}
\caption{{\bf Integration of a hardware decoder into a control system.} (a) Sketch of the \ankaa quantum processing unit (QPU) and control system architecture with the integrated FPGA decoder. Inter-unit latencies are denoted with worst-case values where measured. A (in black): Delay between readout transmit and readout receive (approx. $500$~ns). This delay contributes to overall latency but not to inter-shot or inter-readout delay; B (in yellow): Time required for readout receiver to generate a classified result, convert to intra-unit message, and serialize to the low-latency results crossbar ($120$~ns); C (in yellow): Time required for transmission from readout receiver to low-latency results crossbar ($120$--$160$~ns); D (in black): Time required to handle result message and prepare for broadcast ($40$~ns); E (in teal): Time required to transmit from low-latency results crossbar to internal controller option cards ($120$--$160$~ns), with additional time for readout-to-sequencer logic to send classified results to sequencers, within each card ($40$--$90$~ns); F (in black): Inter-node delay time for broadcasting between control system chassis ($240$--$260$~ns). (b) Mapping of \ankaa sublattice to readout units. Each on-chip multiplexed readout feedline is associated with separate readout-RX and readout-TX gate cards. Because of where these qubits are located on their respective readout feedlines, the sublattice [Q36, Q37, Q38, Q43, Q44, Q45, Q50, Q51, Q52] corresponding to the stability experiment (shown in background and details in \cref{sec:stability}) falls across six different readout TX/RX hardware regions, marked with grey rectangles.}
    \label{fig:control_system_diagram}
\end{figure*}

Quantum computers have the potential to perform computations that are beyond the capabilities of classical computers \cite{shor_algorithms_1994, grover_fast_1996}. However, various quantum algorithms that offer an advantage over classical algorithms will require hundreds of qubits and over a billion operations \cite{gidney_how_2021, reiher_elucidating_2017, blunt_perspective_2022, lee_even_2021}, which are prone to errors. To combat the noise and unlock the potential of such algorithms, quantum error correction (QEC) will be necessary. QEC encodes the information we want to protect by distributing it across multiple qubits and repeatedly generating data characterising quantum errors \cite{terhal_quantum_2015, gottesman_stabilizer_1997, campbell_roads_2017}. Classical algorithms, known as \textit{decoders}, use this data to identify errors that occurred during quantum computation. At low enough physical error rates, QEC suppresses errors when executing quantum algorithms, offering a path towards practical quantum computation. However, QEC codes alone cannot implement a universal and transversal gate set \cite{eastin_restrictions_2009}. Leading proposals for implementing non-Clifford gates require logical branching -- a logical operation conditional on a corrected observable, which is computed by combining the measured value of the observable and the logical correction returned by the decoder. For example, on the surface code \cite{bravyi_quantum_1998, fowler_surface_2012, dennis_topological_2002}, a logical $T$ gate can be implemented via teleportation and magic state injection \cite{bravyi_magic-state_2012, bravyi_universal_2005, litinski_game_2019}, which requires a conditional logical $S$ gate. Since logical branching happens during circuit execution and depends on the decoding result, this places throughput and latency requirements on the decoding process.

The rate at which the decoder processes data (the throughput) needs to be greater than the rate at which data is generated, which on superconducting devices can be as fast as one data extraction round per $1$~\unit{\micro\second}~\cite{battistel_real-time_2023, van_dijk_electronic_2019, jeffrey_fast_2014, acharya_suppressing_2023}. This is necessary to avoid the backlog problem \cite{terhal_quantum_2015}, i.e., an exponential slowdown of the quantum computation due to a growing backlog of data at each decoding iteration. High throughput can be achieved both with fast decoders and by parallelizing the decoding workload over many decoder threads \cite{skoric_parallel_2023,tan2023scalable,bombin2023modular,wu2023fusion,acharya_quantum_2024}. However, parallelization strategies will be limited by \cite{skoric_parallel_2023} a second key parameter:  the \textit{full decoding response time}, which is the time between the final data extraction round and the application of a logical conditional gate. This response time includes the decoding time as well as the communication and control latency times to send the data to and from the decoder. This parameter can be a significant speed bottleneck for operations involving logical branching such as logical non-Clifford gates on surface codes \cite{bravyi_magic-state_2012, bravyi_universal_2005, litinski_game_2019}. As a result, it directly affects the logical clock rate (i.e., the inverse of the time required to perform a single logical non-Clifford gate), another QEC metric that determines the execution speed of fault-tolerant algorithms. 
Reducing both the decoding time and the communication and control latency to minimize the full decoding response time is key to ensure that complex quantum algorithms are executed within reasonable times. For example, when estimating that $2048$-bit RSA integers can be factored in 8 hours using 20 million noisy superconducting qubits, the authors assume a full decoding response time within $10$~\unit{\micro\second} \cite{gidney_how_2021}. To ensure fast decoding response time it is important to decode in \textit{real time}, meaning that data is passed to the decoder and  processed as soon as it becomes available. This contrasts with offline decoding, where the decoding can be performed at any time after the data has been collected. Real-time decoding has been experimentally demonstrated but with decoding response times far exceeding the $10$~\unit{\micro\second} goal \cite{ryan-anderson_realization_2021, acharya_quantum_2024} or using an unscalable lookup table approach \cite{ryan-anderson_realization_2021, egan_fault-tolerant_2021, riste_real-time_2020, bluvstein_logical_2024}.

In this work, we present a real-time decoded QEC experiment on $8$ qubits with logical branching that measures the full decoding response time to be $9.6$~\unit{\micro\second}, including $6.5$~\unit{\micro\second} decoding time and $3.1$~\unit{\micro\second} communication and control latency times, per $9$ measurement rounds. We achieve this by decoding with a scalable FPGA implementation of a Collision Clustering decoder \cite{barber_real-time_2023}, integrated into Rigetti's \ankaaTM superconducting device's control system (\cref{fig:control_system_diagram}). We demonstrate decoding times per QEC cycle faster than the $1$~\unit{\micro\second} threshold for generating measurement data on a superconducting qubit device, ensuring the backlog problem is avoided, and show logical error suppression for a so-called \textit{stability experiment} \cite{gidney_stability_2022} with up to $25$ decoding rounds. The ability of our decoding system to avoid the backlog problem and maintain low-latency feedback opens the door for experiments involving logical branching, which will be vital for implementing a fault-tolerant universal gate set.


\section{Stability experiment on a surface code} \label{sec:stability}
\begin{figure*}[t]
    \centering
    \includegraphics[width=\linewidth]{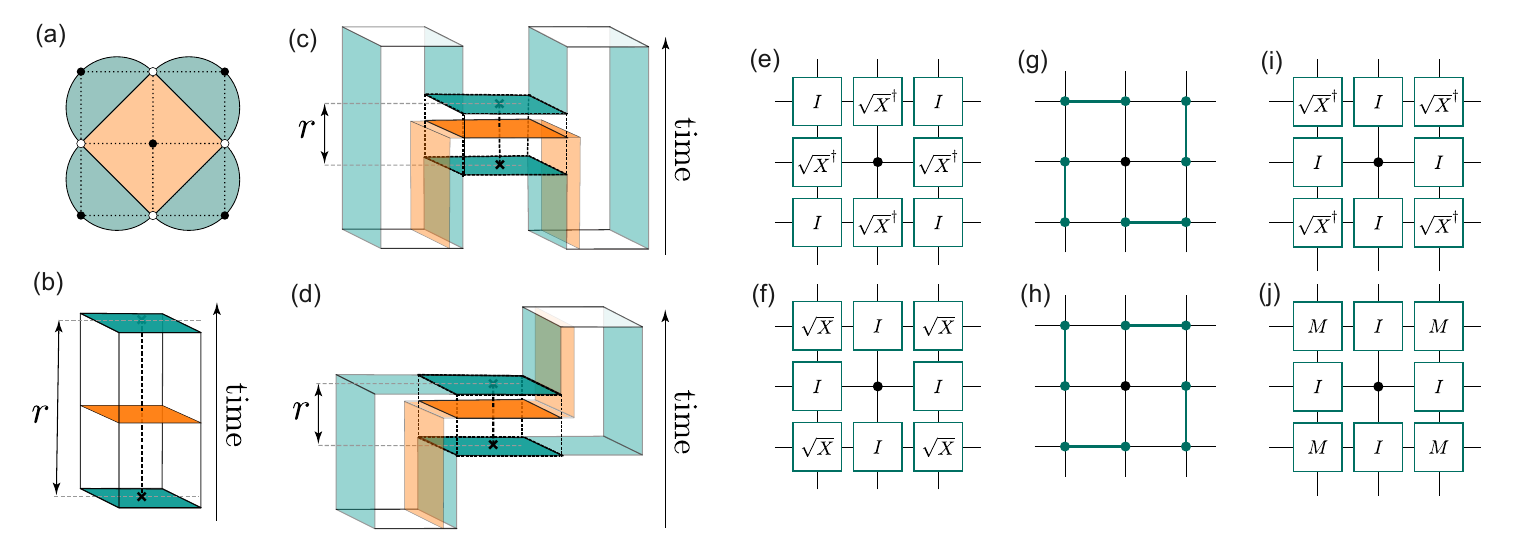}
    \caption{{\bf Stability experiment spacetime diagrams and circuit.} (a) $2 \times 2$ stability patch. White dots correspond to data qubits; black dots correspond to ancilla qubits used to measure the stabilizers. Green coloured half-disks are weight-$2$ $Z$-stabilizers, and the orange square is a weight-$4$ $X$-stabilizer. The dotted lines indicate the hardware connectivity needed to execute a stability experiment. (b--d): Spacetime diagrams for (b) a stability experiment, (c) lattice surgery and (d) logical patch moving. Orange sheets represent locations where errors can flip the value of a logical observable; green sheets represent places where an error can cause an isolated defect, therefore allowing error strings to terminate; the dashed line shows an example string of undetectedable errors that flip the logical observable. The stability experiment is also part of the lattice surgery and logical patch moving routines, as highlighted by the dark sheets in (c) and (d). To preserve the observable in space, it is measured $r$ times. (e--j): Gate layers applied during the $8$-qubit stability experiment, where the operations in (e) prepare the data qubits in the initial state, in (f--i) map to ancilla qubits the weight-$2$ $Z$-stabiliser values and in (j) measure the weight-$2$ $Z$-stabiliser values by measuring the ancilla qubits. The qubit placement matches the one shown in \cref{fig:control_system_diagram}(b). Black edges show the qubit hardware connectivity. Edges overlaid with dark green indicate a $CZ$ gate. Single-qubit gates are depicted as green boxes with labels indicating the applied gates. Measurement and qubit idling are denoted by $M$ and $I$, respectively. The black dot corresponds to the qubit not used in the $8$-qubit stability experiment. Gate layer (e) is only performed at the beginning of the experiment, while layers (f--j) are repeatedly executed and correspond to a single syndrome extraction round. Note that ancilla qubits are not reset between rounds. Gate layers (e), (f) and (i) are executed in $\sim 40$~ns, (g) in $\sim 176$~ns, (h) in $\sim 112$~ns and (j) in $\sim 948$~ns with an additional $336$~ns ring down time. A single QEC round therefore takes approximately $1.7$~\unit{\micro\second}.}
    \label{fig:stability}
\end{figure*}

In this study, we focus on the stability experiment \cite{gidney_stability_2022} implemented with the rotated planar surface code, which is one of the most promising QEC codes due to its high threshold and nearest-neighbour hardware-connectivity requirements \cite{bravyi_quantum_1998, fowler_surface_2012, dennis_topological_2002}. The practical implementation of fault-tolerant gates on the surface code has also been extensively studied and developed \cite{horsman_surface_2012, litinski_game_2019, mcewen_relaxing_2023, fowler_low_2019, fowler_low-overhead_2012}. To realize a QEC protocol, Pauli operators called \textit{stabilizers} are repeatedly measured to detect physical errors. For the rotated surface code, the stabilizer set consists of weight $2$ or $4$ tensor products of $X$ or $Z$ operators arranged on a 2D square lattice of qubits. A \textit{detector} is a product of stabilizer measurements, whose value is deterministic in the case of no errors \cite{gidney2021stim}. For example, we typically assign a detector to the product of consecutive measurements of the same stabilizer. A detector whose value has been flipped from the one expected in the case of no errors is referred to as a \textit{defect}. Therefore, defects signal the presence of errors. The \textit{defect rate} is the probability of observing a defect. A \textit{syndrome} is the set of values of detectors observed during an experiment.

A stability experiment \cite{gidney_stability_2022} tests the capability of a quantum processing unit (QPU) to preserve an observable in space by correctly measuring products of stabilizers (\cref{fig:stability}(a,b)). Surface code operations such as lattice surgery \cite{horsman_surface_2012, erhard_entangling_2021} (\cref{fig:stability}(c)), logical qubit patch movement \cite{litinski_game_2019} (\cref{fig:stability}(d)) or the logical Hadamard gate \cite{bombin_logical_2023, geher_error-corrected_2024} all involve measuring products of stabilizers. In these examples, incorrectly measuring the stabilizer product can introduce a logical error. For instance, in lattice surgery, this stabilizer product determines the logical branching decisions when performing non-Clifford gates, with a logical error meaning the wrong branch is followed. The stability experiment \cite{gidney_stability_2022} verifies the ability to protect against these logical errors and in this work acts as a smaller-scale simulation of logical branching experiments. Such an experiment on a surface code-like patch is performed with an over-complete set of stabilizers whose product is equal to the identity, yet not assigned as a detector. We will refer to this product of stabilizers as a \textit{logical observable}  \cite{gidney2021stim}. In the absence of errors, the logical observable is necessarily measured as $+1$. This means that we can verify the ability of a QEC scheme to correctly deduce whether the logical observable has been flipped. For a stability experiment at low enough physical error rates, the logical error probability is suppressed as the number of decoding rounds is increased.

In this work, we focus on real-time decoding of a $2\times2$ stability experiment only measuring an overcomplete set of 4 stabilizers of $Z \otimes Z$ form, and omitting the single $X \otimes X \otimes X \otimes X$ stabilizer that plays no role in decoding. This omission simplifies the syndrome extraction circuit and leads to improved logical error rates. We refer to this experiment as the \textit{stability-$8$} experiment. \cref{fig:stability}(e) shows the data qubit preparation at the start of the experiment. To obtain a single round of syndrome data, we use the syndrome extraction circuit shown in \cref{fig:stability}(f--j), where steps (f--i) map the product of stabilizers to the ancilla qubits, which are then measured in step (j). The measurement outcomes are aggregated to pre-defined detector outcomes, which are inputs to the decoder.


\section{Integrating an FPGA decoder in the \ankaa control system} \label{sec:ankaa_device}

We decode with an FPGA decoder integrated into the \ankaa control system (more details on the integration work are given in Methods). \ankaa is a superconducting transmon qubit device, with $84$ qubits arranged on a square lattice, and two-qubit gates implemented via tunable couplers \cite{manenti_full_2021, sete_error_2024, sete_parametric-resonance_2021} (see Methods for details on the \ankaa device). The decoder we use is an FPGA implementation of a Collision Clustering decoder \cite{barber_real-time_2023}, which achieves the same logical fidelities as the Union Find algorithm \cite{delfosse_almost-linear_2021}. Modifications were made to the FPGA decoder of Ref.~\cite{barber_real-time_2023} to allow for decoding the stability experiment and syndrome extraction circuits without mid-circuit reset operations \cite{geher_reset_2024, ali_reducing_2024, krinner_realizing_2022}. The FPGA decoder is optimised for speed and scalability to larger systems. The decoder executes at $156.25$~MHz frequency on one of the sequencers responsible for gate pulses on the \ankaa control system (see Methods for more details). It has also been verified for operating at $400$~MHz frequency, allowing room for further speed improvements compared to those demonstrated here.

The communication between the FPGA decoder and the rest of the control system is performed via custom assembly language-level instructions. We built a prototype compiler that receives a pyQuil \cite{smith2016practical} program and adds the necessary instructions to perform real-time decoding and store the results (see Supplementary Information, Section D for more details). Our compiler can also merge two pyQuil programs into a logical branching experiment, where the second program is executed conditionally based on a real-time decoded measurement outcome obtained during the first program.

The stabilizer measurement outcomes are automatically passed to the decoder during circuit execution. Once the decoder receives the allocated number of measurements, the measurements are converted to a syndrome and decoding is performed. The decoded result, along with data storing the decoder speed and status, is written to allocated registers. In a logical branching experiment, the FPGA decoder's result is communicated in real time to gate sequencers which then, conditionally on the received result, execute the second program.


\section{Logical error probability suppression}

\begin{figure*}[t]
    \centering
    \includegraphics[width=\linewidth]{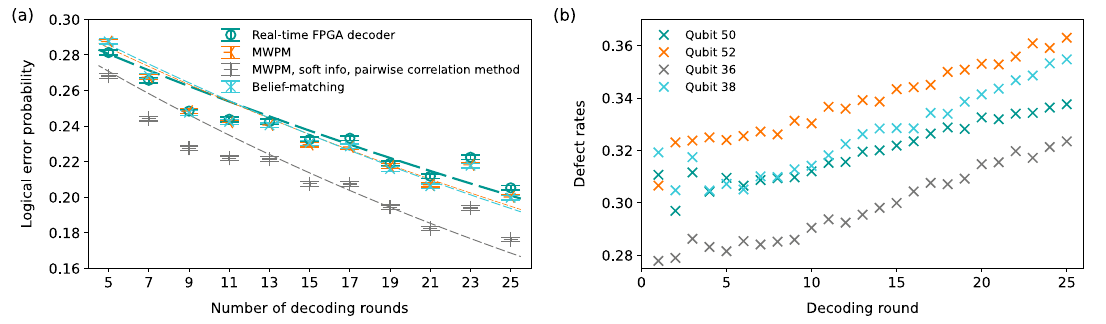}
    \includegraphics[width=\linewidth]{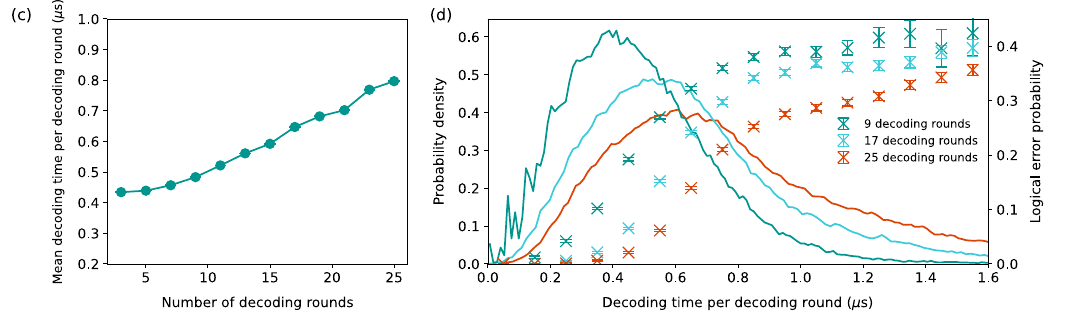}
    \caption{{\bf Logical error probabilities and decoder timings.} (a) Logical error probabilities for the stability-$8$ experiment executed on the \ankaa device as a function of number of decoding rounds. We plot logical error probabilities calculated using the decoding results of the real-time FPGA decoder as well as the following offline decoders -- minimum-weight perfect matching (MWPM), belief-matching and a variation of MWPM that uses soft information and the pairwise correlation method to construct the decoding graph. Error bars show the standard error of the mean. Dashed lines are included to guide the eye. (b) Defect rates as a function of decoding round for each of the four ancilla qubits, for a stability-$8$ experiment performed with $25$ decoding rounds. Qubit IDs correspond to the layout shown in \cref{fig:control_system_diagram}(b). (c) Mean decoding time per decoding round for the stability-$8$ experiment, decoded using the real-time FPGA decoder operating at $156.25$~MHz. Error bars are computed as the standard error of the mean and are smaller than the graph point size. (d) Left axis, solid lines: Distributions of decoding times per decoding round for stability-$8$ experiments performed with $9, 17$ and $25$ decoding rounds. Right axis, crosses: Logical error probabilities as a function of decoding time. Repetitions of the stability experiment are placed according to their decoding time in bins of width $100$~ns, and each logical error probability is estimated as an average within a bin. Error bars are computed as the standard error of the mean.}
    \label{fig:half_stab_speed}
\end{figure*}

We perform stability-8 experiments on the \ankaa device for varying numbers of QEC rounds, corresponding to $5$ to $25$ decoding rounds, and decode in real time. We repeat each experiment $10^5$ times to obtain the probability of a logical error persisting at the end of the QEC routine. In \cref{fig:half_stab_speed}(a), we show that on \ankaa using real-time FPGA decoding the physical error rates are low enough to demonstrate logical error probability suppression with an increasing number of decoding rounds, which is the signature of a successful stability experiment. The logical error probabilities decrease from $(28.1 \pm 0.1)\%$ at $5$ decoding rounds to $(20.5 \pm 0.1)\%$ at $25$ decoding rounds. We also decode offline with minimum-weight perfect matching (MWPM) \cite{higgott_sparse_2023} and belief-matching \cite{higgott_improved_2023} decoders (see Methods for more details) and show in \cref{fig:half_stab_speed}(a) that when using hard measurement data (i.e., the readout signal classified as ``0'' or ``1''), the real-time FPGA decoder results are comparable to these decoders. We also explore decoding with MWPM decoder using soft measurement information (i.e., the in-phase and quadrature components of the raw readout signal) \cite{pattison2021improved} and a decoding graph constructed with the pairwise correlation method \cite{chen_calibrated_2022, spitz2018adaptive, chen_exponential_2021, ali_reducing_2024, pattison2021improved} (see Supplementary Information, Section C). This decoder achieves the lowest logical error probability, $(17.6 \pm 0.1)\%$, as using soft information and pairwise correlation methods gives the decoder more detailed information, resulting in higher accuracy decoding. In the future, we expect FPGA decoders to be adaptable so that soft information with the pairwise correlation method could also be used with FPGA decoding.

We observe fluctuations in the logical error probability estimates with varying numbers of decoding rounds, relative to the expected decay; this is likely due to fluctuations in device performance over time which have been previously shown to occur in superconducting circuit devices, such as fluctuations in the coherence times due to two-level systems (TLSs) \cite{tls_martinis_2018, tls_poletto_2015}. \cref{fig:half_stab_speed}(b) shows a slight increase in defect rates as further measurement rounds are performed, most likely due to a combination of leakage and heating effects. We believe the limitations on the logical error probability reported here to be largely due to the mid-circuit readout fidelity of the ancilla qubits and the fidelity of the $CZ$ gates that were available on the device; the \ankaa processor design was not optimised for these operations in particular. For a more detailed discussion of the device design and relevant error channels, see Methods and Supplementary Information, Section A.


\section{Decoding throughput}

It is necessary to ensure the decoder has a high enough throughput to avoid the backlog problem \cite{terhal_quantum_2015, skoric_parallel_2023}, which arises when data is generated at a faster rate than it can be decoded. If this is not the case, then, because each conditional operation needs to wait for the ever-growing backlog of data to be decoded before it can be applied, each subsequent conditional operation will be exponentially slower, eventually halting the computation. On the \ankaa device, a single round of syndrome extraction takes approximately $1.7$~\unit{\micro\second} (see \cref{fig:stability}(f--j)). Other superconducting devices have demonstrated a single round of syndrome extraction of approximately $1$ \unit{\micro\second} \cite{battistel_real-time_2023, van_dijk_electronic_2019, jeffrey_fast_2014, acharya_suppressing_2023}. \cref{fig:half_stab_speed}(c) shows that our mean decoding time per round ranges from $0.44$~\unit{\micro\second} when decoding $5$ rounds to $0.79$~\unit{\micro\second} when decoding $25$ rounds. These values remain below the $1$ \unit{\micro\second} threshold for data generation on a superconducting qubit device, providing strong evidence that the backlog problem will be avoided when operated as a streaming decoder \cite{skoric_parallel_2023, bombin2023modular, tan2023scalable}. The increase in decoding time per round with more decoding rounds is due to the higher noise in later rounds, leading to increased defect rates in deeper circuits (\cref{fig:half_stab_speed}(b)). As the device improves and the heating and leakage effects are reduced, we expect the mean decoding time per round to grow more slowly as the number of decoding rounds increases. Simulations of the FPGA decoder speed show that the decoding times increase superlinearly with number of decoding rounds, but it is known the scaling can be improved with further optimizations \cite{barber_real-time_2023}. In the future, we also expect the FPGA decoder to be parallelised, further reducing decoding times.

In \cref{fig:half_stab_speed}(d), we examine the distribution of the decoding times per round (solid lines) and group experiment repetitions in $100$~ns bins, based on their decoding time per round, and plot these against the average logical error probability of the experiments in the bins (crosses). We observe that most of our decoding times are well below $1$~\unit{\micro\second} per decoding round. Such experiments correspond to a high-likelihood of correctly decoding the syndrome. A small proportion of experiment repetitions exceed the $1$~\unit{\micro\second} threshold, most likely due to a higher number of defects affecting those experimental runs, which slows down the completion of the decoding. The data represented by crosses indicate that such experiments are associated with worse logical error probabilities.

For practical fault-tolerant computation, we will need lower physical error rates to achieve much lower logical error rates. Lower physical error will lead to lower defect rates and therefore significantly reduce the decoding times. There is also room to more than double the speed of decoding reported in this work, since the FPGA decoder has been verified to be able to operate at $400$~MHz, instead of the $156.25$~MHz frequency used in these experiments.


\section{Control and communication latencies} \label{sec:fast_feedback}

\begin{figure}
    \centering
    \includegraphics[width=0.85\linewidth]{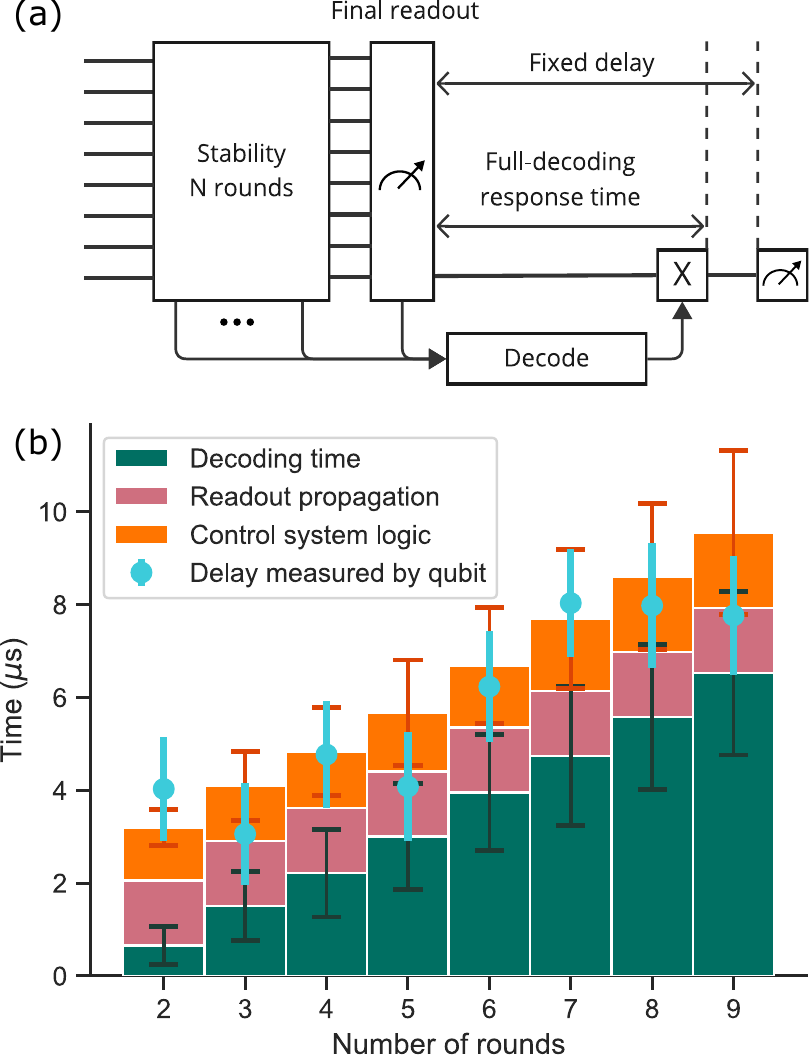}
    \caption{{\bf Full decoding response time experiment.} (a) The experiment consists of a number of rounds of the stability experiment. After the final readout, the data is sent to the decoder, and a conditional $X$ gate is applied to one of the qubits as soon as decoding result is available in the control system. The qubit is measured after some fixed delay. (b) Full decoding response time measurements as a function of number of rounds. Total time measured by the control system FPGA clock cycles (bars) consists of the decoding time measured by decoder onboard metric units (green), waiting for all readouts to reach the sequencer with the decoder (pink), as well as additional logic such as constructing syndrome packets and sending them to the decoder, receiving the result and applying the conditional gate (orange). This is compared to the delay measured from the qubit T1 time (blue points) in order to assess any additional delays in the wiring that might be unaccounted for by the control system (for details see Supplementary Information, Section E). Error bars on bars are the standard deviation of the decoder times distribution (green) and the total control system time (orange) while the error bars on the blue points are the standard error of the mean of the delay measured by the qubit.}
    \label{fig:feedback_timing}
\end{figure}

We now focus on the full decoding response time, to quantify the contributions of the communication and control latency in our system. We design an experiment that serves as a stepping stone to logical conditional gates. 
We perform a circuit that conditionally applies a physical gate based on the outcome of decoding the stability-$8$ experiment (see \cref{fig:feedback_timing}(a)). After the QEC experiment completes and the decoder has received the allocated number of measurement outcomes of the stability circuit, we start the real-time FPGA decoder and wait for the result. When the result is received, we conditionally apply an $X$ gate on the qubit controlled by the gate sequencer on the same FPGA as the decoder. The qubit is read out after a fixed delay, set longer than the worst recorded decoding time (approx. $N_{\mathrm{rounds}}$ $\mu s$ for an $N_{\mathrm{rounds}}$ experiment).

We measure the full decoding response time as a function of the number of QEC rounds (see \cref{fig:feedback_timing}(b)). Firstly, we record the FPGA clock cycles on the control system sequencer and the cycles on the decoder accelerator. We note that while the majority of the full decoding response time can be attributed to the decoding time for a larger number of rounds, there is a significant additional latency incurred by the control system delays.  In particular, $1.4$~\unit{\micro\second} is the time for the final readout results to reach the decoder, with the remaining $1$-$1.5$~\unit{\micro\second} ($250-370$ FPGA clock cycles) consisting of additional control system logic: collecting the measurements into packets to be sent to the decoder, sending them via the WISHBONE bus (see \cref{sec:integration_details}), receiving results and performing the conditional logic (see Supplementary Information, Section D). Furthermore, since the error bars for the timing of the decoder and full control system logic closely match in size, we conclude that the decoding time accounts for the majority of the variance in the full decoding response time.

To check that there are no additional unaccounted delays, we also measure the full decoding response time by using the qubit T1 time as a clock. The measurement distribution after the fixed delay (see \cref{fig:feedback_timing}(a)) conditional on the final QEC round readout is a function of the full decoding response time, measurement fidelities, and the qubit's T1 time (see Supplementary Information, Section E for the detailed derivation). Thus, by collecting the relevant reference data immediately prior to running the feedback experiment, we can estimate the delay experienced by the qubit (see \cref{fig:feedback_timing}(b)) which takes into account any delays that might not be measured by the control system clock. With this, we confirm that there are no additional significant delays as the estimated full decoding response time closely matches the one measured by the control system clock.

For a $9$-decoding-round experiment, we find the full decoding response time to be $9.6$~\unit{\micro\second}, including $6.5$~\unit{\micro\second} decoding time and $3.1$~\unit{\micro\second} communication and control latencies.
Thanks to the FPGA implementation of our fast decoder and its integration into the \ankaa system, in the small-distance studies reported in this work we keep the full decoding response time within the order of $d$ \unit{\micro\second}, where $d$ is the number of decoding rounds. Maintaining this condition will be crucial when scaling up $d$, as it will ensure that this response time will not be a critical limiting factor for the logical clock speed when implementing non-Clifford operations \cite{skoric_parallel_2023,bombin2023modular}.

\section{Discussion}\label{sec:discussion}

Real-time decoding had been previously demonstrated using lookup table decoders on trapped-ion quantum computers \cite{ryan-anderson_realization_2021, egan_fault-tolerant_2021}, superconducting qubits \cite{riste_real-time_2020} and a neutral atom device \cite{bluvstein_logical_2024}. The memory requirements for lookup table decoders scale exponentially with the number of qubits in the code, meaning that this approach is not feasible for practical fault tolerance, which will require larger codes \cite{litinski_game_2019, reiher_elucidating_2017, blunt_perspective_2022, lee_even_2021}. The Collision Clustering decoder implementation has been designed to be memory-efficient, with simulations demonstrating that even for distance-$23$ memory experiments, the decoder uses less than $13$~KB of memory \cite{barber_real-time_2023}.

Recently, the team at Google demonstrated real-time decoding for the distance-$5$ memory experiment \cite{acharya_quantum_2024}, utilising $54$ qubits. In their experiment, the measurement data is sent via Ethernet cable from their control system to a software decoder. They demonstrate high throughput, avoiding the backlog problem, by parallelizing decoding \cite{skoric_parallel_2023, bombin2023modular, tan2023scalable}. The reported decoding response time includes an average of $63$~\unit{\micro\second} to process the final $10$ QEC cycle block. They require an additional $10$~\unit{\micro\second} communication latency to input data to the decoder, and further additional contributions that are not reported including feedback communication latency. Overall, these numbers are significantly slower than our full decoding response time of $9.6$~\unit{\micro\second} for a $9$ decoding rounds experiment. However, we caution against a direct comparison given the significantly different experimental conditions in which our team's and Google's experiments were conducted. Our experiment utilises fewer qubits, while operating at a higher noise regime. These conditions have opposite effects on the decoding time, as the decoding time decreases with smaller code sizes but increases with higher noise in the device. The performance of our FPGA decoder has been extrapolated via simulations to show that, at a noise just below threshold for memory experiments ($p=0.5\%$), the FPGA decoder will be able to maintain a decoding time below $1$~\unit{\micro\second} per round for memory experiments with distances up to $11$ \cite{barber_real-time_2023}. This, combined with the significant reduction of control and communication latencies achieved by integrating the decoder in the QPU's control system, suggests that our full decoding response time will remain sufficiently fast to demonstrate larger-scale QEC experiments \cite{gidney_how_2021}. We also suggest several strategies for further reducing latency times as the code size is increased. Firstly, readout propagation can be reduced by having a more direct line of communication between readout sequencers and the decoding stack through a standardized interface. Secondly, reordering and packing classified measurements before sending them to the decoder can take up to ${\sim} 400$~ns at $9$ rounds. In the future, we foresee that such commonly used functionality will have to be offloaded to custom hardware logic attached to the decoder. With these, we do not expect the control system latencies to be a major contributor to the full decoding response time.

\section{Conclusion}\label{sec:conclusion}

The experimental results presented demonstrate an implementation of a decoding protocol that enables next generation QEC experiments requiring logical branching. We report mean decoding times per round under the $1$~\unit{\micro\second} threshold, avoiding the backlog problem. The fast-feedback experiment shows that integrating an FPGA decoder into the \ankaa control system ensures a low-latency full decoding response time of $9.6$~\unit{\micro\second} with a total of $9$ decoding rounds for an $8$-qubit experiment.

We show that the logical error probabilities are suppressed in a stability-8 experiment as the number of decoding rounds increases. However, practical fault tolerance will require lower error rates than those reported here. Fine-tuning the readout pulses to be faster and higher fidelity will be essential to achieving this. While resets are not available it will also be beneficial to tune the readout pulses to better preserve the classified state of the qubits. Using resets should improve the results \cite{geher_reset_2024} (see exploration of this in Supplementary Information, Section F). Another primary focus should be to increase the ratio of qubit coherence to two-qubit gate duration, along with reducing leakage and other coherent errors.

In the future, fault-tolerant quantum algorithms that have the potential to outperform classical algorithms will require codes utilizing a large number of operations \cite{gidney_how_2021, reiher_elucidating_2017, litinski_game_2019, blunt_perspective_2022, lee_even_2021}. To support this, the FPGA decoder should be updated to be a streaming decoder \cite{skoric_parallel_2023, bombin2023modular, tan2023scalable}. To improve the accuracy of decoding results, we expect future FPGA implementations to enable updates to the decoding graph in real time, allowing for soft-information, leakage-aware decoding \cite{suchara_leakage_2015} or correlated decoding \cite{fowler_optimal_2013}. Combining such advances to decoding and improvements to the device with the methods developed in this work will pave the way for logical operations such as magic state teleportation, enabling fault-tolerant quantum algorithms.


\begin{acknowledgements}
We thank  Steve Brierley, Nicolas Didier, Rossy Nguyen, Matthew Reagor, David Rivas, Jake Taylor, Alice Voaden, and Catriona Wright for their support and championing of this project.

We thank Maria Maragkou and Luigi Martiradonna for feedback on the manuscript.

We thank David Byfield and Mark Turner for advice on implementation and use of analysis tools.
\end{acknowledgements}

\section*{Author contributions}

L.C., L.S., N.S.B. and A.D.H. led the direction of the project with E.T.C. providing guidance and advice. L.C., L.S., N.S.B. and A.R. implemented the software infrastructure to set up experiments and compile programs for real-time decoding and implemented analysis tools. L.C., L.S., N.S.B. and A.R. ran the experiments and analysed the results. J.A.V and A.D.P. fine-tuned and analysed Ankaa-2 performance and ran experiments. E.T.C., O.C., G.P.G., H.K., E.M., N.I.G. assisted with the theoretical aspects and aided in interpreting the results of the stability experiment. K.J., K.M.B., T.B., O.B., C.T., L.S. contributed to the technical implementation of the FPGA decoder. J.Mc. and G.J. integrated the FPGA decoder into the Ankaa-2 control system. S.P. provided guidance during technical pathfinding. K.S. and M.B. authored novel translation stack features, enabling low-level hardware control through Rigetti’s production stack. A.V.G. implemented resets. J.Ma. implemented soft information decoding. L.C., L.S., N.S.B. and E.T.C. drafted the manuscript and designed the figures with contributions to Methods from J.Mc., J.A.V., A.D.P., A.D.H. and to Supplementary Information from J.A.V., A.D.P., J.Ma., A.R. and A.V.G. Comments and edits were incorporated from all authors. All authors read and approved the final manuscript. 

\section*{Competing interests}

The authors declare no competing interests.

\section*{Additional information}

Correspondence and requests for materials should be addressed to L.C., L.S. and N.S.B.


\bibliography{main}


\section*{Methods} \label{sec:methods}

\subsection{Description of control system} \label{sec:control_system}

The Rigetti control system is built around a configurable card cage chassis. Each physical card in a chassis has inputs or outputs targeting specific RF bands and features high-speed digital-to-analog (DAC) and analog-to-digital (ADC) converters that interface with a Kintex-7 FPGA. The FPGA designs generate (or receive) control signals using one or more \textit{sequencers}, each of which consist of a dedicated processor, waveform scheduler, and waveform generator. Experiments created in pyQuil are compiled into program binaries, consisting of waveforms and instructions in a proprietary assembly language, which are loaded into each relevant sequencer’s memory by the control system before execution.

The assembly language allows for communication with external modules implemented in the FPGA design via instructions that interact with I/O ports. Additionally, the sequencer assembly language includes instructions which load classified results into a sequencer’s memory so that it can be used by a program. 

In order to support experiments that require real-time results, the control system features a low-latency network that distributes each qubit’s most recently classified result when it is published to this network, as part of the readout capture pipeline. Onboard-classified data propagates through the control system with a worst case latency time that is calculated beforehand and used when constructing programs to ensure valid results have been received. 

\subsection{Real-time FPGA decoder integration} \label{sec:integration_details}
When integrating the FPGA decoder and control system, we selected the FPGA design with the lowest resource utilization, the gate drive card design, which is responsible for generating microwave pulses to enact single-qubit gates. We used a version of the FPGA decoder featuring a 32-bit WISHBONE interface for communication. We integrated the decoder into the design by adding I/O ports to the gate drive pulse sequencer that initiate READ and WRITE clock cycles over the WISHBONE interface. Via these I/O ports, a sequencer program was able to read from, and write to, any of the decoder’s addresses in order to perform actions such as writing the syndrome register, starting the decoding process, and reading the decoder result. Signals from other ports on the decoder core, such as the decoder’s status bitfield, were also made available to sequencer programs via an I/O port.

An experiment using an $8$-qubit system requires more than one control system chassis. We selected one chassis to be the hub of a star network and connected all other relevant chassis to it, ensuring the lowest achievable results propagation time by limiting the inter-chassis communication to a single hop. Interactions with a decoder were handled by a gate drive sequencer on the hub chassis, which had access to the classified results from the other chassis.

While the sequencer processor clock operates at $250$~MHz, we opted to drive the FPGA decoder using the $156.25$~MHz clock already present in the design to simplify the timing requirements. We added clock-crossing logic to enable I/O port instructions coming from the sequencer clock domain to result in WISHBONE transactions in the decoder clock domain. Signals received directly from ports on the decoder (e.g., the decoder status bits) were also transitioned to the sequencer clock domain using clock-crossing logic. In the future we intend to use a single clock domain for the sequencer processor and decoder.

\subsection{\ankaa device specifications} \label{subsec:ankaa_2_specifics}

The \ankaa device is a superconducting circuit composed of 84 qubits and 149 tunable couplers, all of which are composed of statically capacitively coupled tunable transmons. Each qubit transmon is also coupled to a coplanar waveguide resonator for readout. The device is arranged in a square array with 7 columns and 12 rows of qubits, with a tunable coupler between each neighboring qubit pair, two readout feed lines coupled to each column (in groups of 6), charge drive lines coupled to each individual qubit, and flux lines coupled to the SQUID loop within each individual transmon.

The selected sublattice used when executing the stability-8 experiments was composed of qubits 36, 37, 38, 43, 45, 50, 51, and 52. Qubits 36, 38, 50 and 52 were used as the ancilla qubits, and had a median measurement fidelity of 93.3\%. The median readout fidelity across the entire \ankaa device was 94.7\% at this time but, as the ancilla qubits were repeatedly measured during the experiment, they required optimisation for mid-circuit readout (discussed in Supplementary Information, Section A), which is the main cause of their reported readout fidelity being below the device median for the values reported here.  The single-qubit gates across the sublattice had a median isolated randomized benchmarking fidelity of 99.86\%, and the $CZ$s had a median interleaved randomized benchmarking fidelity of 97.3\% (see Supplementary Information, Section A for a summary of the associated error budget). The median two-qubit gate length was $104$~ns, and measurement pulse length was $964$~\unit{\nano\second} with additional $336$~ns ring down time. The sublattice had a median T1 of $13.2$~\unit{\micro\second} and a median T2 of $10.6$~\unit{\micro\second}.

\subsection{Software decoding} \label{subsec:sw_decoding}

In addition to decoding in real time with the FPGA decoder, we also obtain decoding results on the same data with offline MWPM \cite{higgott_sparse_2023} and belief-matching \cite{higgott_improved_2023} decoders. We use implementations in \textit{PyMatching~2} \cite{higgott_sparse_2023} and \textit{BeliefMatching} \cite{higgott_improved_2023} for the MWPM and belief-matching decoders, respectively. We use Stim \cite{gidney2021stim} to set up the circuits that represent our experiments and use these circuits to initialise the software decoders. For belief-matching the belief propagation stage uses the product-sum method with maximum of $5$ iterations. Note that in our experiments we do not observe any improvement when decoding with belief-matching. This is because our decoding graph does not contain hyperedges, as we do not measure the middle stabilizer in the stability-$8$ experiment. Belief-matching tends to be beneficial when hyperedges are present in the decoding graph.

\clearpage

\end{document}


\title{Supplementary Information for ``Demonstrating real-time and low-latency quantum error correction with superconducting qubits''}
\author{Laura Caune}
\email{laura.caune@riverlane.com}
\affiliation{Riverlane, Cambridge, CB2 3BZ, UK}
\author{Luka Skoric}
\email{luka.skoric@riverlane.com}
\affiliation{Riverlane, Cambridge, CB2 3BZ, UK}
\author{Nick S. Blunt}
\email{nick.blunt@riverlane.com}
\author{Archibald Ruban}
\affiliation{Riverlane, Cambridge, CB2 3BZ, UK}
\author{Jimmy McDaniel}
\affiliation{Rigetti Computing, 775 Heinz Avenue, Berkeley, California 94710, USA}
\author{Joseph A. Valery}
\affiliation{Rigetti Computing, 775 Heinz Avenue, Berkeley, California 94710, USA}
\author{Andrew D. Patterson}
\affiliation{Rigetti UK Ltd, 138 Holborn, London, EC1N 2SW, UK}
\author{Alexander V. Gramolin}
\affiliation{Riverlane, Cambridge, Massachusetts 02142, USA}
\author{Joonas Majaniemi}
\affiliation{Riverlane, Cambridge, CB2 3BZ, UK}
\author{Kenton M. Barnes}
\affiliation{Riverlane, Cambridge, CB2 3BZ, UK}
\author{Tomasz Bialas}
\affiliation{Riverlane, Cambridge, CB2 3BZ, UK}
\author{Okan Bu\u{g}daycı}
\affiliation{Riverlane, Cambridge, CB2 3BZ, UK}
\author{Ophelia Crawford}
\affiliation{Riverlane, Cambridge, CB2 3BZ, UK}
\author{Gy\"{o}rgy P. Geh\'{e}r}
\affiliation{Riverlane, Cambridge, CB2 3BZ, UK}
\author{Hari Krovi}
\affiliation{Riverlane, Cambridge, Massachusetts 02142, USA}
\author{Elisha Matekole}
\affiliation{Riverlane, Cambridge, Massachusetts 02142, USA}
\author{Canberk Topal}
\affiliation{Riverlane, Cambridge, CB2 3BZ, UK}
\author{Stefano Poletto}
\affiliation{Rigetti Computing, 775 Heinz Avenue, Berkeley, California 94710, USA}
\author{Michael Bryant}
\affiliation{Rigetti Computing, 775 Heinz Avenue, Berkeley, California 94710, USA}
\author{Kalan Snyder}
\affiliation{Rigetti Computing, 775 Heinz Avenue, Berkeley, California 94710, USA}
\author{Neil I. Gillespie}
\affiliation{Riverlane, Cambridge, CB2 3BZ, UK}
\author{Glenn Jones}
\affiliation{Rigetti Computing, 775 Heinz Avenue, Berkeley, California 94710, USA}
\author{Kauser Johar}
\affiliation{Riverlane, Cambridge, CB2 3BZ, UK}
\author{Earl T. Campbell}
\affiliation{Riverlane, Cambridge, CB2 3BZ, UK}
\affiliation{Department of Physics and Astronomy, University of Sheffield, UK}
\author{Alexander D. Hill}
\affiliation{Rigetti Computing, 775 Heinz Avenue, Berkeley, California 94710, USA}

\date{\today}

\thispagestyle{plain}
\pagestyle{plain}

\maketitle

\tableofcontents

\appendix

\section{Targeted calibration of \ankaa for the stability circuits} \label{si:ankaa_2_calibration}

\begin{figure}
    \centering
    \includegraphics[width=1\linewidth]{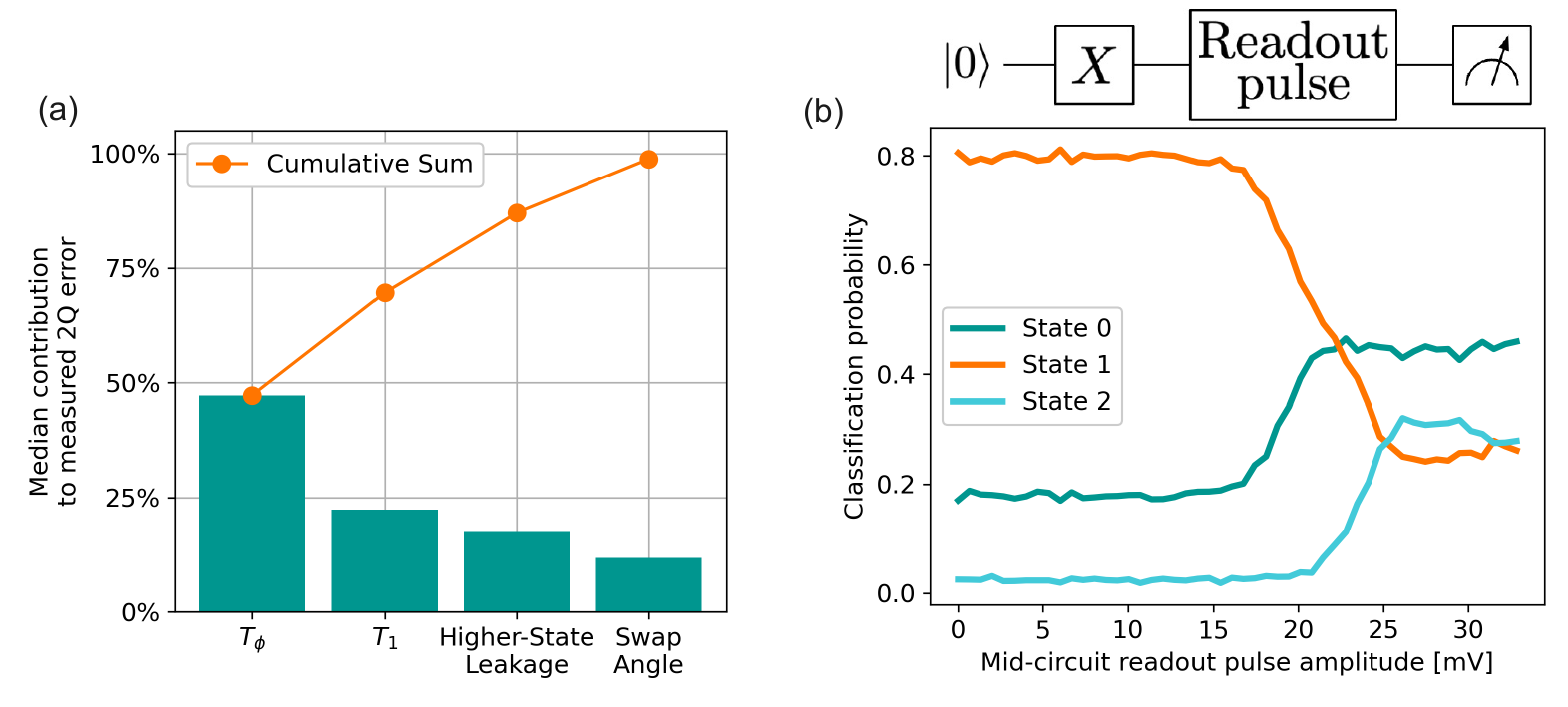} 
    \caption{{\bf Error budget for $CZ$ gates and measurement-induced leakage experiment.} (a) $T_\phi$ and $T_1$ errors are calculated using coherence times measured at both the gate operating point and the qubit idling points (relevant during the short, idle padding delays that play around each gate) \cite{sete_error_2024}. Higher-state leakage error is calculated via the leakage randomized benchmarking protocol \cite{wood_leakage_2018}, using three-state readout to analyze the decay of population in the computational subspace over the course of an interleaved RB experiment. Swap angle error is error attributable to off-target values of $\theta$ in the fSIM unitary defined in \cite{arute_floquet_2020} -- for an ideal $CZ$, $\theta=0$. Ref.~\cite{arute_floquet_2020} also describes a series of Floquet calibration circuits, the first of which is used here to measure $\theta$ for these $CZ$ gates. Considering medians taken over the lattice, $T_\phi$ contributes almost half of the measured error, with the incoherent error comprised of $T_\phi$ and $T_1$ together representing approximately $70\%$. All together, these four error channels account for a median $99\%$ of measured error for these $CZ$ gates. (b) Representative example of experimental results detailing the dependence of measurement-induced leakage on readout pulse power, shown here for Qubit 52. The circuit consists of an initial preparation of the target qubit in the excited state, before transmission of a test readout pulse, followed by a sufficiently long delay for the resonator to be depopulated before a final measurement with a pulse previously and separately optimized for three-state classification at the end of a circuit. The test readout pulse is the pulse under study, with the process aiming to allow it to be improved for mid-circuit use in stability experiments. The resulting population of the target transmon in each of its three lowest energy states at the end of the circuit is plotted here as a function of the amplitude of the test readout pulse. At low amplitude, population remains largely in the first excited state, as desired, up to a threshold limited by the classifier's three-state confusion matrix. However, there exists an amplitude threshold above which the apparent population rapidly begins to more evenly distribute across the three states. This phenomenon resembles the numerical results presented in \cite{shillito_ionization_2022}, which describes the process by which strong resonator drives can suddenly excite coupled transmon transitions to states above their Josephson potential wells. For each ancilla qubit used in this study's stability circuits, these threshold amplitudes were measured and used as upper bounds in further parallel readout optimizations in order to ensure that each mid-circuit measurement would avoid this error mechanism.}
    \label{fig:cz_and_readout}
\end{figure}

The design of the \ankaa circuit Hamiltonian is optimized towards the application of $iSWAP$ gates.  Reconfiguration of the steady state qubit frequencies was required across a sublattice to support the $CZ$ gates used in this work. We compute that the primary source of $CZ$ errors comes from $T_\phi$ errors, with additional contributions from $T_1$ errors, higher-state leakage, and swap angle errors (see \cref{fig:cz_and_readout}(a)). The selected sublattice used in the experiments (qubits 36, 37, 38, 43, 45, 50, 51 and 52) was selected based on the ability to enable the necessary resonances between |02⟩ (or |20⟩) and |11⟩ across the contained qubit pairs, and the performance of the sublattice while in such a configuration. Measurement operations on this chosen sublattice were specifically optimized towards the mid-circuit use case. For example, the readout pulses that were optimised for maximum classification fidelity when positioned at the end of a circuit were observed to cause a large accumulation of leakage to the second-excited state of the repeatedly measured transmons when running the stability circuits over multiple rounds. Reducing the amplitude of readout pulses below an experimentally deduced threshold resulted in a large increase in the probability that the state of the measured qubit was preserved in the detected state after the readout process (see \cref{fig:cz_and_readout}(b)). Other considerations included shortening the duration of the readout process by approximately a factor of two compared to standard circuit-end readout, as well as specifically optimizing the ancilla qubit readout pulse parameters for maximum fidelity when performed simultaneously.

\section{Decoding graphs} \label{si:decoding_details}

\begin{figure*}[ht]
\includegraphics[width=0.5\columnwidth]{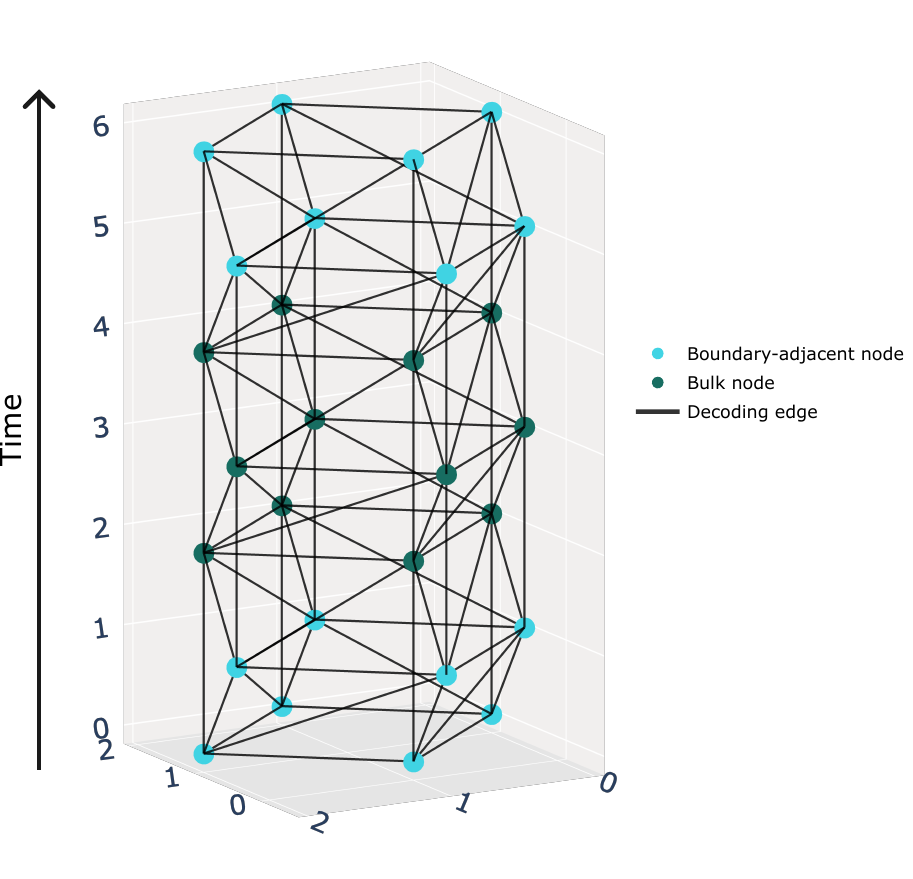}
\caption{\label{fig:decoding_graph}
{\bf Stability experiment decoding graph.} An example of the decoding graph for 8 rounds of syndrome measurements (equivalent to 7 rounds of detectors). The odd layers are slightly offset so that the edges connecting layers two rounds apart are visible. These edges arise since we are not using mid-circuit resets. The diagonal edges between layers are due to the errors that can happen in the middle of the syndrome extraction circuit and are known as ``hook'' errors \cite{dennis_topological_2002}.  Boundary-adjacent nodes are detectors that can be matched with a virtual boundary node, and in stability experiments such boundaries are at the top and the bottom rather than on the sides.}
\end{figure*}

\cref{fig:decoding_graph} shows the decoding graph used by the real-time FPGA decoder and software decoders to decode the stability-$8$ experiment in this work. The FPGA decoder decoding graph does not use edge weights. For the software decoders that do not use the pairwise correlation method, we use a circuit-level noise model to obtain the edge weights for the decoding graph. Since we did not build a noise model tailored for \ankaa, we opted to use a standard noise model used to model the noise in superconducting qubit devices. This model captures the fact that on superconducting qubit devices the two qubit gates and measurements are typically noisier than single qubit gates. The noise model we use is the same as in Ref.~\cite{barber_real-time_2023}, from which we quote the noise channels applied for a fixed probability $p=0.03$:

\begin{quote}
\begin{itemize}
    \item Depolarisation of both qubits after each two-qubit gate with probability $p$.
    \item Depolarisation of each idle qubit and after each single-qubit gate, including measurement and reset operations, with probability $p/10$.
    \item Measurement flip with probability $p$. 
\end{itemize}
\end{quote}

We decided to use $p=0.03$ based on the median two qubit gate fidelities on \ankaa.

\section{Soft information decoding} \label{si:soft_info}

In the the main text we report the lowest logical error probability when using MWPM software decoder that takes a decoding graph from the pairwise correlation method and is dynamically updated with soft measurement information. Here we describe how the soft measurement information is provided to this decoder.

The minimum-weight perfect matching (MWPM) decoder \cite{higgott2023sparse} operates on a graph representation of the decoding problem, where detectors correspond to nodes, and possible error mechanisms correspond to edges. The probability of an error mechanism $p_e$ is mapped to the weight of an edge via $w(e)=-\log(p_e/(1-p_e))$, meaning that high-likelihood errors have a low-weight edge in the decoding graph. By finding the lowest-weight perfect matching in the graph, the MWPM algorithm effectively approximates the most probable errors that caused the observed defects. We set the weights in the graph according to the pairwise correlation method \cite{spitz2018adaptive, chen2022calibrated}, which infers the probabilities of error mechanisms based on the defect frequency in the experimental data. To further improve the accuracy of the decoder, we use rich measurement information (soft information) from the qubit readout \cite{pattison2021improved, ali_reducing_2024} to dynamically update the probabilities of the measurement error mechanisms in the decoding graph.

Soft information here refers to the integrated voltage $z=(I, Q)$ from the measurement response of the readout resonator, where $I$ and $Q$ are the in-phase and in-quadrature components of the complex signal, respectively. The soft outcome $z$ is passed to a measurement classifier and converted to an outcome probability $P(\hat{z} \mid z)$ corresponding to the possible measurement outcomes $\hat{z}\in \{0, 1\}$. The hard measurement outcomes are given by $\hat{z}=\text{argmax}{[P(0 \mid z), P(1 \mid z)]}$.

To train the classifier, we perform a calibration run where we prepare and measure each transmon qubit in the $\ket{0}$- and $\ket{1}$-states $5\times 10^4$ times. We leverage a linear discriminant classifier from the Python library \texttt{scikit-learn v1.3.2} \cite{scikit-learn} as our classifier, and use it to predict the measurement outcome probabilities $P(0 \mid z)$ and $P(1 \mid z)$ for each soft measurement $z$ in the experiment. By preparing an equal number of initial states $\ket{0}$ and $\ket{1}$ for the training data set of the classifier, we ensure the prior probabilities of the two outcomes are $P(0)=P(1)=1/2$. The outcome probabilities are used to update the edge weights in the decoding graph according to 

\begin{equation}\label{eq:soft-weights}
    w(z) = -\log\left[\frac{P(z \mid 1 - \hat{z})}{P(z \mid \hat{z})}\right],
\end{equation}
where we use Bayes' theorem to get $P(z \mid \hat{z}) = P(\hat{z} \mid z)P(z)/P(\hat{z})$. We then use the MWPM decoder to decode the hardened syndrome on the updated decoding graph to obtain our logical corrections.  The results of this decoder, as discussed in the main text, show a large improvement in the logical fidelity relative to other decoders that use hard measurement data.

\section{Detailed experiment workflow with real-time decoding and fast-feedback} \label{si:experiment_workflow}

\begin{figure}[!hb]
\includegraphics[width=0.5\linewidth]{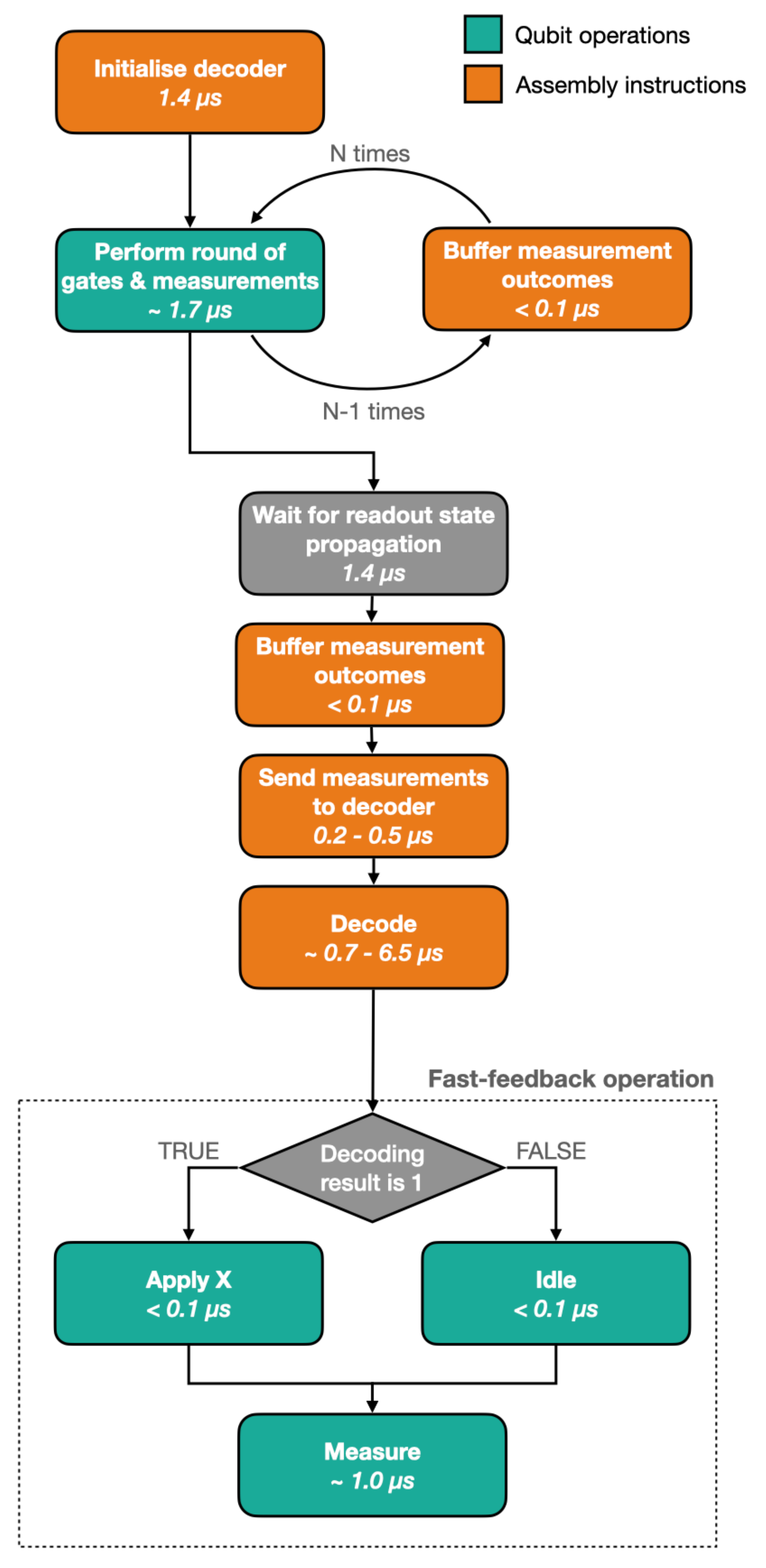}
\caption{\label{fig:experiment_flowchart}{\bf Flowchart for the real-time decoded experiment with fast-feedback.} The execution time for each element is shown in microseconds. When this depends on the number of rounds in the experiment, the times for $2$ and $9$ measurement rounds are shown instead.}
\end{figure}

The FPGA decoder is integrated into the \ankaa control system. Tasks specific to decoding in real time and with fast feedback are executed by interleaving the qubit operations of the stability-8 experiment with instructions written in the control system's proprietary assembly language, as shown in \cref{fig:experiment_flowchart}. Such tasks include management of the measurement outcomes, communication with the decoder, and carrying out the fast-feedback operation.

\noindent The main experiment components are:
\begin{enumerate}
    \item \textbf{Initialise decoder.}
Writes to the decoder experiment features (e.g. number of rounds) and decoding configurations (e.g. logical observable definition).

    \item\textbf{Perform round of gates \& measurements.}
Executes a single QEC round of the stability-8 experiment. All rounds are identical besides the last in which the data qubits are also measured. Mid-circuit measurements are followed by ring-down delays (included in measurement times), which allow the resonators to decay back to their ground state before the start of the following round \cite{bengtsson_model_2024}.

    \item\textbf{Buffer measurement outcomes.}
Stores the outcomes of the latest measurement round in the decoder sequencer's memory. Even though buffering takes less than $0.1$~\unit{\micro\second}, a $1.4$~\unit{\micro\second} delay must take place between measurement and buffering to allow for the measurements to propagate from the readout sequencers to the decoder sequencer. To address this, buffering is swapped with the following round, enabling the readout state to propagate in parallel with each QEC round and preventing the qubits idling mid-circuit. This instruction swapping is applied to all mid-circuit rounds, ensuring that only the last round experiences an actual readout state propagation delay. Note that while \cref{fig:experiment_flowchart} does not reflect this instruction reordering, the overall duration it shows for a given number of rounds is accurate.

\item\textbf{Send measurements to the decoder.}
Collects, formats and writes all measurement outcomes to the decoder. This consists of pushing each round's buffer to the sequencer's data stack, combining them into a series of $32$-bit binary strings and writing them sequentially to the decoder.

    \item\textbf{Decode.}
The decoder computes the syndrome from measurement outcomes and decodes it using the Collision Clustering algorithm \cite{barber_real-time_2023}. A write instruction to the decoder initiates decoding, followed by a polling of the decoder's status register, which stalls the program until decoding completes. The decoding result is a Boolean describing whether the decoder computed that the physical errors flipped the stability experiment's logical observable.

    \item\textbf{Execute fast-feedback operation.}
Applies an $X$ gate conditionally on the decoding result and measures the qubit. The $X$ gate is applied if the result is $1$, otherwise the qubit is left to idle until measurement for a time equal to the gate's duration.
\end{enumerate}

\section{Full decoding response time experiment}
\label{si:fast_feedback}
\begin{figure*}[t]
    \includegraphics[width=\linewidth]{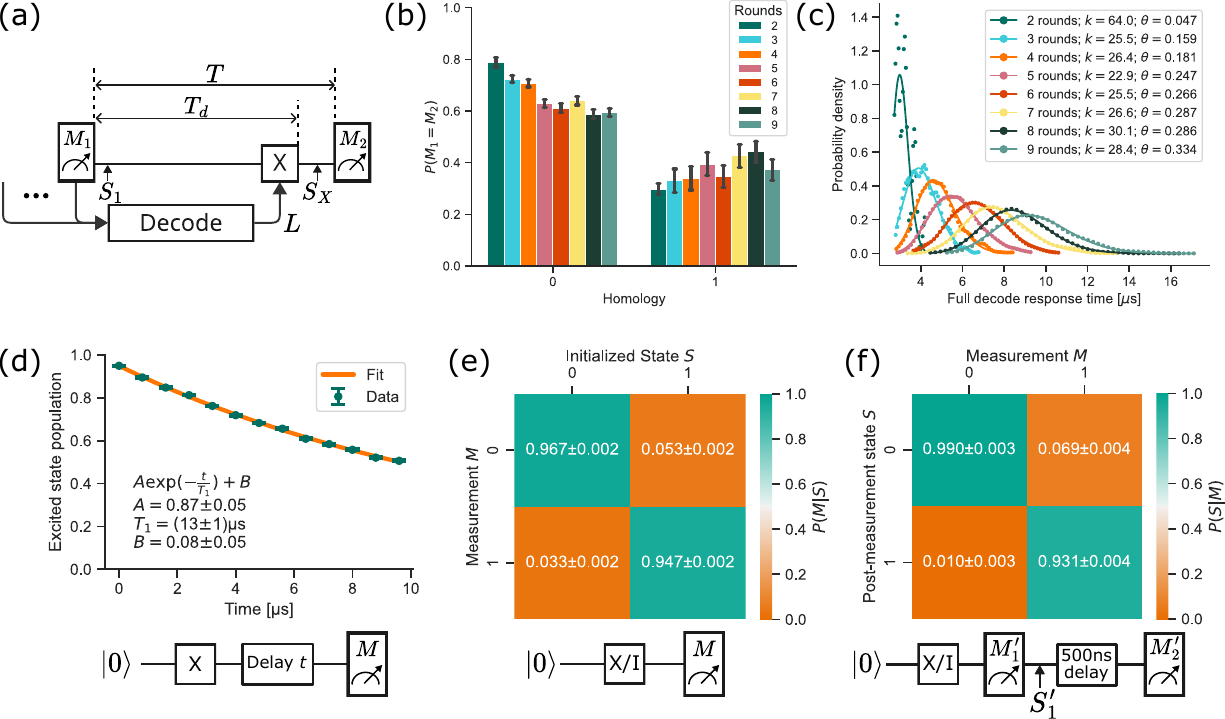}
    \caption{{\bf Full decoding response time experiment timing parameters.} (a) The circuit used for timing the full decoding response time. After the final measurement of the stability circuit $M_1$, the qubit is in state $S_1$ that is evolving freely while waiting for the conditional gate to be applied. Depending on the logical correction $L$, an $X$ gate is applied $T_d$ time after $M_1$, changing the qubit state to $S_X$. A fixed time $T$ after $M_1$ a second measurement $M_2$ is performed.
    (b) Measured probability of the two measurements matching ($M_1 = M_2$) depending on the computed logical correction $L$ for different number of rounds of the stability experiment. The error bars are standard errors of the mean.
    (c) Distribution of the full decoding response times as measured by the control system with fits corresponding to the probability density function of Gamma distribution $\Gamma (k, \theta)$ where $k, \theta$ are the usual shape and scale parameters respectively.
    (d--f) Reference data used to compute $T_d$. The circuits used to acquire the data are shown below the plots. (d) $T_1$ qubit relaxation time. The data is fitted with the exponential decay curve and error bars are standard deviations of the mean. (e) Measurement confusion matrix $\pr{(M=i|S=j)}\ i,j=0,1$ . (f) Post-measurement state given the measurement result  $\pr{(S'_1=i|M'_2=j)}\ i,j=0,1$ acquired by measuring the state twice, with a 500 ns ring-down time between measurements.}
    \label{fig:fast_feedback_reference}
\end{figure*}

As described in the main text, in addition to measuring the response time in the control system, we also measure the delay to applying the conditional operation by considering its effect on the qubit $T_1$ decay. This experiment allows us to perform full end-to-end response timing purely based on the qubit's physics. In the main text the results of this experiment were acquired on a different day from the data measuring logical error probabilities and throughput leading to a difference in total decoding times due to the variations in device performance and therefore defect rates. The relevant part of the circuit and parameters are shown in \cref{fig:fast_feedback_reference}(a).

First, we verify that indeed the conditional operation is applied correctly. We expect that if the decoding result $L$ is $L=0$ then the first measurement $M_1$ is equal to the second measurement $M_2$; and if $L=1$ then the first measurement $M_1$ is opposite to the second measurement $M_2$. We can see that is indeed the case in \cref{fig:fast_feedback_reference}(b), although the distribution has significant noise due to multiple microseconds of qubit idling between the measurements.

Second, we derive three reference datasets to classify the qubit's $T_1$ time decay and measurement characteristics (\cref{fig:fast_feedback_reference}(d--f)). All reference data is acquired immediately prior to running the feedback experiment. To simplify the analysis, we assume that state initialisation and single-qubit operations are perfect as their fidelity is much higher than the measurement fidelity.
We calculate the qubit's $T_1$ decay by initializing the $\ket{1}$ state and measuring after a fixed delay $t$ (see \cref{fig:fast_feedback_reference}(d)). By fitting an exponential decay, we find
\begin{align}
    \label{eq:t1}
    \pr(M=1 | S=1, t) = m(t) = A\exp{(-t/T_1)} + B\\
    A = 0.87 \pm 0.05;\ T_1 = (13 \pm 1) \, \mu \mathrm{s};\ B=0.08 \pm 0.05. \nonumber
\end{align}
We further evaluate the measurement confusion matrix, which we label $\pr(M=i | S=j) = P^{M|S}_{ij}$ by preparing a qubit in state $\ket{S=j}$ and then measuring, obtaining the outcome $M=i$ (see \cref{fig:fast_feedback_reference}(e)).

Next, we calculate the distribution of states immediately after the measurement $M'_1$ given the measurement result $\pr(S=i | M=j) = Q^{S|M}_{ij}$ by performing a second measurement $M'_2$  after a $500$ ns ring down (see \cref{fig:fast_feedback_reference}(f)). Here, we use the prime symbol to distinguish these measurements from the ones in \cref{fig:fast_feedback_reference}(a). Considering the probability of the second measurement conditioned on the first, the law of total probability and the Markovian property $P(M'_2=i|M'_1=j,S'_1=k) = P(M'_2=i|S'_1=k)$ gives us
\begin{align}
    \label{eq:state_given_mmt}
    \pr(M'_2=i|M'_1=j) & = \sum_{k=0,1} \pr(M'_2=i|S'_1=k) \pr(S'_1=k|M'_1=j) , \\ \nonumber
    & = \sum_{k=0,1}  P^{M|S}_{ik} Q^{S|M}_{kj},
\end{align}
where $S'_1$ is a state post-$M'_1$. Since the $\ket{0}$ state is approximately stable and the $\ket{1}$ state decays with a known lifetime, we assume
\begin{equation}
    \label{eq:state_given_mmt_cases}
\pr(M'_2=1|S'_1=k)=
        \begin{cases}
            P^{M|S}_{10} & \text{if } k=0, \\
            m(t_r) & \text{if } k=1,
        \end{cases}
\end{equation}
with $t_r=500$ ns being the ring-down delay and $m(\cdot)$ is the T1 decay fit defined in \cref{eq:t1}. Using these, \cref{eq:state_given_mmt} becomes a set of linear equations for $Q^{S|M}_{ij}$ that we solve to acquire the matrix in \cref{fig:fast_feedback_reference}(f). All uncertainties are propagated using Python uncertainties package \cite{uncertainties}.

Having the reference data, we go back to the fast-feedback experiment (\cref{fig:fast_feedback_reference}(a)). 
We consider the probability of measuring $M_2=1$ given the decoding result (i.e. logical correction) $\pr(M_2=1 | L=i)$. Using the law of total probability, we expand in state $S_1$ after the first measurement to give  
\begin{align}
    \pr(M_2=1 | L=i) &= \pr(M_2=1 | L=i, S_1 = 0) \pr(S_1 = 0) + \pr(M_2=1 | L=i, S_1=1) \pr(S_1 = 1) .
\label{eq:probm2}
\end{align}

We can re-express the terms in \cref{eq:probm2} in terms of the reference data $P^{M|S}_{ij}, Q^{S|M}_{ij}$ and $m(t)$. The probability to have $S_1=i$ can be deduced from the $M_1$ distribution and $Q^{S|M}_{ij}$ as
\begin{equation}
\label{eq:state}
    \pr(S_1 = i) = \sum_{j=0,1} Q^{S|M}_{ij} \pr(M_1 = j).
\end{equation}
Assuming that the state that starts at $S_1=0$ stays at 0 during $T_d$ and gets flipped to 1 by the $X$ gate if $L=1$, we have
\begin{equation}
\label{eq:probm2_cases}
\pr(M_2=1 | L=i, S_1 = 0) = 
        \begin{cases}
            P^{M|S}_{10} & \text{if }L=0, \\
            m(T-T_d) & \text{if }L=1,
        \end{cases}
\end{equation}
where $m(\cdot)$ is the function defined in \cref{eq:t1}.

Next, we determine the total time $T$ between the measurements $M_1$ and $M_2$ (see \cref{fig:fast_feedback_reference}(a)). $T$ consists of the readout propagation delay, control system logic and an additional delay of $\sim N_{\mathrm{rounds}}\ \mu$s inserted on the measurement sequencer at the moment the decoder starts decoding. This can be calculated by timing the program execution on the control system, but here we decide to calculate it directly from the acquired distributions to ensure we are capturing any unaccounted delays. We do this by post-selecting for $L=0$ cases where there is no conditional operation so the experiment resembles the one in \cref{fig:fast_feedback_reference}(f). We have $ \pr(M_2=1 | L=0, S_1=1) = m(T)$ and thus, from \cref{eq:probm2},
\begin{align}
    \pr(M_2=1 | L=0) = P^{M|S}_{10} \pr(S_1 = 0) + m(T) \pr(S_1 = 1), \nonumber \\
    m(T) = \frac{\pr(M_2=1 | L=0) - P^{M|S}_{10} \pr(S_1 = 0)}{\pr(S_1 = 1)},
\label{eq:probm2_h0}
\end{align}
where we can replace $\pr(S_1)$ using \cref{eq:state} so the right hand is entirely in term of measurable quantities. Taking the inverse of $m$ gives us $T$. Finally, knowing $T$, we can compute $T_d$ by considering the distribution in \cref{eq:probm2} in the case when the conditional operation is applied ($L=1$). Using \cref{eq:probm2_cases}, we have $\pr(M_2=1 | L=1, S_1 = 0) = m(T-T_d)$, and we use the law of total probability, conditioning on the post-$X$ gate state $S_X$ to obtain
\begin{align}
    \pr(M_2=1 | L=1, S_1=1) &= \pr(M_2=1|S_X=1)\pr(S_X=1|S_1=1, L=1) + \pr(M_2=1|S_X=0)\pr(S_X=0|S_1=1, L=1) \nonumber \\
                     &=m(T-T_d)\left(1-e^{-T_d/T_1}\right) + P^{M|S}_{10}e^{-T_d/T_1}.
\end{align}
Note that $m(T-T_d) = [m(T) - B]e^{T_d/T_1} + B$, so setting $\alpha_d=e^{-T_d/T_1}$ and putting everything together in \cref{eq:probm2}, we find
\begin{align}
    \pr(M_2=1 | L=1) &= \left(\frac{m(T) - B}{\alpha_d} + B\right)\pr(S_1=0) + \left[\left(\frac{m(T) - B}{\alpha_d} + B\right)(1-\alpha_d) + P^{M|S}_{10}\alpha_d\right]\pr(S_1=1),
\end{align}
which is a quadratic equation for $\alpha_d$ in terms of measurable quantities and quantities derived in  \cref{eq:t1,eq:state,eq:probm2_h0}. Note that the quantity evaluated by putting in the experimental data is $\langle\alpha_d\rangle = \ee \left(e^{-T_d/T_1}\right)$ where $T_d$ is a random variable. Computing $\Tilde{T}_d = -T_1 \log \ee (e^{-T_d/T_1})$ gives us a biased estimate for $\ee (T_d)$. In the following, we show that this bias is negligible given the distribution of decoding times, allowing us to estimate $\ee (T_d) \approx -T_1 \log \langle\alpha_d\rangle$.

We model $T_d$ as distributed by the Gamma distribution $T_d \sim \Gamma(k, \theta)$ where $k, \theta$ are the shape and scale parameters respectively. The probability density function is defined as:
\begin{equation}
    f(x; k, \theta) = \frac{x^{k-1}e^{-x/\theta}}{\theta^k\Gamma(k)}.
\end{equation}
In \cref{fig:fast_feedback_reference}(c) we see that this distribution fits well the full decoding response times as measured by the control system clock, and we have no reason that this would stop being true for the full response time as experienced by the qubit. By the law of the unconscious statistician (LOTUS):
\begin{align}
    \ee \left(e^{-T_d/T_1}\right) &= \int_0^{\infty} e^{-x/T_1} \frac{x^{k-1}e^{-x/\theta}}{\theta^k\Gamma(k)} dx \nonumber \\
    \label{eq:etd_integral}
    &= \frac{1}{\theta^k\left(1/\theta + 1/T_1\right)^k} \int_0^{\infty} \left(\frac{1}{T_1} + \frac{1}{\theta}\right)^k\frac{x^{k-1}e^{-x\left(\frac{1}{T_1} + \frac{1}{\theta}\right)}}{\Gamma(k)} dx \\
    &= \left( 1 + \frac{\theta}{T_1} \right)^{-k},
    \label{eq:etd_final}
\end{align}
where the integral in \cref{eq:etd_integral} is 1 as it is integrating the probability density function of a Gamma distribution $\Gamma\left(k, \left(1/T_1 + 1/\theta\right)^{-1}\right)$ across the full domain. Hence, using \cref{eq:etd_final}:
\begin{equation}
    \Tilde{T_d} = -T_1 \log \ee \left(e^{-T_d/T_1}\right) = T_1 k \log \left( 1 + \frac{\theta}{T_1} \right).
\end{equation}
If $\theta \ll T_1$, we can expand the logarithm to first order in Taylor series and use the expected value of the Gamma distribution $\ee (T_d) = k\theta$ to find:
\begin{equation}
    \Tilde{T_d} \approx T_1 k \left[ \frac{\theta}{T_1} + \mathcal{O}\left(\frac{\theta^2}{T_1^2}\right)\right] = \ee (T_d) \left[ 1 + \mathcal{O}\left(\frac{\theta}{T_1}\right)\right].
\end{equation}
As we can see from \cref{fig:fast_feedback_reference}(c), the scale parameter for the range of rounds used in this experiment is $\theta < 0.4$. With this, $\theta / T_1 \lesssim 3\%$ is a negligible bias in the estimator as the relative standard error is greater than $10\%$. Even if there are additional delays unaccounted for in the distributions in \cref{fig:fast_feedback_reference}(c), we do not expect these to change $\theta$ significantly enough to invalidate this analysis.

\section{Unconditional qubit resets} \label{subsec:resets}

While the experiments described in the main text did not involve resetting ancilla qubits between QEC rounds, we have also explored and describe here a specific implementation of unconditional qubit resets. Note that these resets were implemented during the final stage of the project and we did not evaluate their performance with the real-time FPGA decoder. Here we present evidence of the improved logical error rates when using unconditional resets with a software decoder.

Recall that qubits can be reset to their ground state either by passively waiting (to achieve a thermal equilibrium with the cold bath, which happens on a time scale much longer than the relaxation time $T_1$) or by actively applying additional control pulses, which may or may not be conditional on the qubit state. Conditional (or measurement-based) resets work by first measuring the qubit and then applying an $X$ gate if the measurement outcome is ``1''. In contrast, unconditional resets apply the same pulse sequence regardless of the qubit's state. While conditional resets cannot improve the performance of QEC circuits, fast and high-fidelity unconditional resets offer a significant advantage by doubling the number of measurement errors that can be tolerated by a stability experiment~\cite{geher_reset_2024}.

Unconditional resets work by transferring the excited-state population of the qubit to a lossy environment, such as the qubit's readout resonator~\cite{geerlings_demonstrating_2013, zhou_rapid_2021}. Although the transferring rate is limited by the qubit-resonator coupling strength $g$, the resonator can be quickly thermalized with the environment if it is strongly coupled to the cold bath: $1 / \kappa \ll T_1$, where $\kappa$ is the dissipation rate of the resonator. This requirement is well satisfied for the \ankaa qubits, where the typical values are $T_1 \approx 15~\mu\text{s}$ and $\kappa / 2\pi \approx 3~\text{MHz}$.

In this work, we investigated the double-drive reset of population (DDROP) protocol~\cite{geerlings_demonstrating_2013}, which is particularly well suited to the \ankaa control system. Compared to the parametric reset protocol of Ref.~\cite{zhou_rapid_2021}, the DDROP scheme requires much smaller instantaneous bandwidth of the control electronics. This is because both the qubit and the resonator drives used for DDROP are detuned from their respective resonant frequencies by only several dozen MHz. The DDROP protocol is also applicable to fixed-frequency qubits, which is another advantage over the parametric resets requiring flux-tunable qubits.

Let us now explain how the DDROP protocol works. It exploits the qubit-resonator coupling, which is described by the well-known Jaynes--Cummings model~\cite{krantz_guide_2019}. Transmon qubits typically operate in the dispersive regime, when the qubit-resonator coupling is weak compared to their detuning: $g \ll |\Delta|$, where $\Delta \equiv \omega_q - \omega_r$ and $\omega_q$/$\omega_r$ is the bare frequency of the qubit/resonator. In this regime, the Jaynes--Cummings Hamiltonian can be written as 
\begin{align}
\hat H &= \hbar \left(\omega_r + \chi \hat \sigma_z\right) \hat a^{\dagger} \hat a + \frac{\hbar}{2} \left(\omega_q + \chi\right) \hat \sigma_z \nonumber \\
&= \hbar \omega_r \hat a^{\dagger} \hat a + \frac{\hbar}{2} \left(\omega_q + \chi + 2\chi \hat a^{\dagger} \hat a\right) \hat \sigma_z, \label{eq:JC}
\end{align}
where $\hbar$ is Planck's constant, $\hat \sigma_z$ is the Pauli $Z$ operator of the qubit, $\hat a^{\dagger}$ and $\hat a$ are, respectively, the creation and the annihilation operators of the resonator (which is modeled as a harmonic oscillator), and $\chi = g^2 / \Delta$ is the so-called dispersive shift (in our case, $\chi < 0$ since $\Delta < 0$).

Two effects caused by the qubit-resonator interaction are immediately apparent from this Hamiltonian. As follows from the first line of Eq.~(\ref{eq:JC}), the resonator frequency $\omega_r$ acquires a shift of $\pm \chi$, whose sign depends on the qubit's state (i.e., the $\pm 1$ eigenvalue of the $\hat \sigma_z$ operator). This effect is widely used for performing a dispersive readout of the qubit by probing the frequency of its readout resonator~\cite{krantz_guide_2019}. The second line in Eq.~(\ref{eq:JC}) implies that the qubit frequency $\omega_q$ also experiences two frequency shifts: a constant shift $\chi$ (called the Lamb shift) and a variable shift $2\chi n$ (known as the ac-Stark shift), whose size depends on the number of photons, $n = \langle \hat a^{\dagger} \hat a \rangle$, populating the resonator.

In the dispersive regime, driving the readout resonator at a frequency $\omega_r + \delta_r$ (we will refer to $\delta_r$ as the resonator detuning) leads to a steady state with the average number
\begin{equation}
\bar n_{g,e} (\delta_r) \approx \frac{n_0}{1 + 4 \left(\delta_r \pm \chi\right)^2 / \kappa^2} \label{eq:n_bar}
\end{equation}
of photons in the resonator, where $n_0$ is a constant proportional to the drive's power~\cite{tosi_effects_2024}. This equation assumes that the qubit is in a specific state: either ground, $|g\rangle$, or excited, $|e\rangle$. For each value of the resonator detuning, the photon number~(\ref{eq:n_bar}) takes two values: $\bar n_g$ if the qubit is in $|g\rangle$ (which corresponds to the ``$+$'' sign in the denominator) and $\bar n_e$ if the qubit is in $|e\rangle$ (the ``$-$'' sign in the denominator). Equation~(\ref{eq:n_bar}) describes two Lorentzian peaks, which are centered at $\pm \chi$ and have the amplitude $n_0$ and the full width at half maximum (FWHM) $\kappa$.

\begin{figure}
\includegraphics[width=\linewidth]{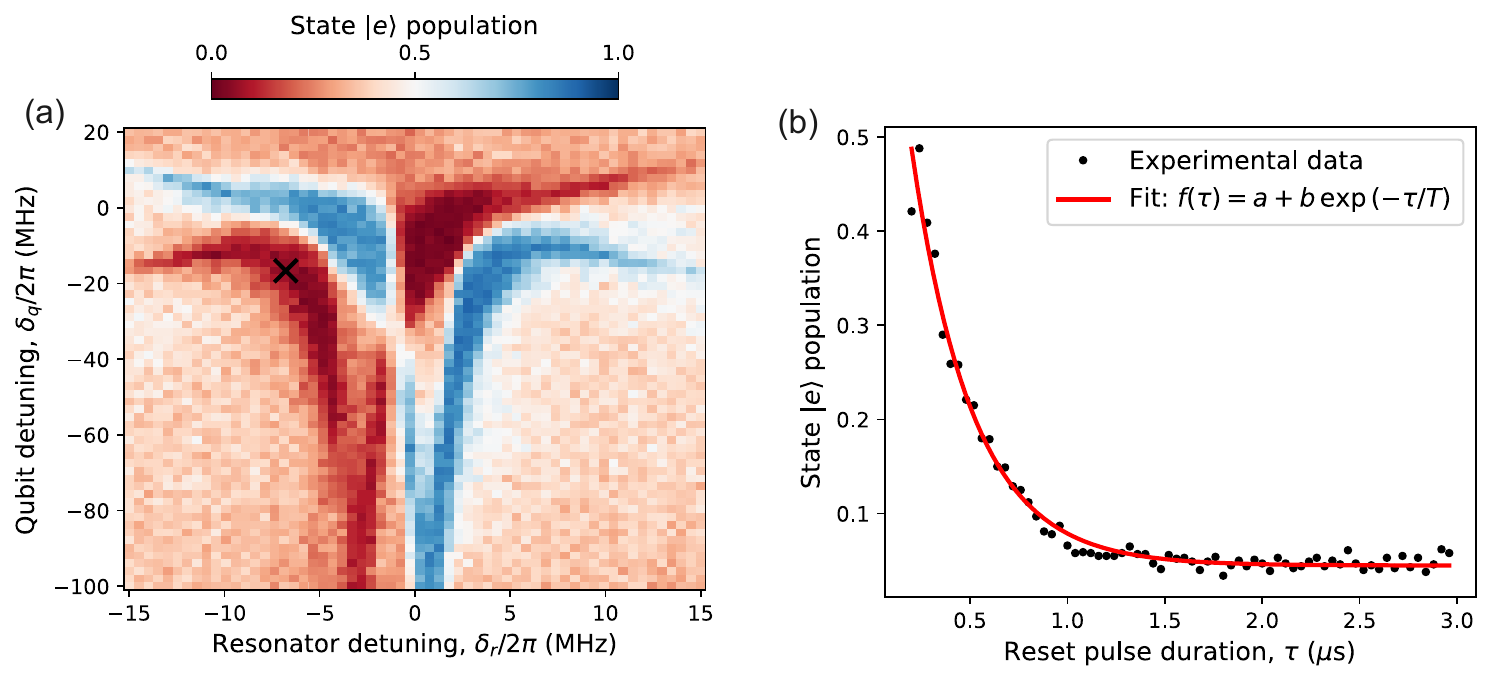}
\caption{\textbf{Calibration of DDROP resets for Qubit~76 of \ankaa.} (a) Results of two-tone spectroscopy. The dependence of the qubit detuning $\delta_q$ on the resonator detuning $\delta_r$ approximately follows Eq.~(\ref{eq:delta_q}). The color represents the measured population of the qubit's state $|e\rangle$ averaged over 200 shots. Two distorted Lorentzian dips can be clearly seen: the left one corresponds to the photon number $\bar n_e$ (red color), while the right one corresponds to $\bar n_g$ (blue color). The separation between the two dips is about $2\chi / (2\pi) \approx -5~\text{MHz}$ and the FWHM of each Lorentzian is $\kappa / (2\pi) \approx 3~\text{MHz}$. The black cross indicates the combination of drive detunings used for implementing the unconditional reset pulse of panel (b). (b) Excited-state population as a function of  the reset pulse duration. Experimental data averaged over 1000 shots are shown with black markers. The red curve is a fit to the data with a function $f(\tau) = a + b \exp{(-\tau / T)}$. We obtained the following best-fit parameters: steady-state population $a = 0.045$ and decay constant $T = 0.31~\mu\text{s}$.}
\label{fig:reset_calibration}
\end{figure}

Due to the Stark effect, the resonator being populated with $\bar n_{g,e}$ photons causes the qubit frequency to shift by
\begin{equation}
\delta_q \approx 2\chi \bar n_{g,e} (\delta_r) \label{eq:delta_q}
\end{equation}
relative to its bare value $\omega_q$ (we call $\delta_q$ the qubit detuning). Therefore, we expect that if we drive the resonator at different frequencies and simultaneously probe the qubit's transition frequency, the latter would be shifted by $\delta_q$ and the observed dependence of $\delta_q$ on $\delta_r$ would approximately follow Eqs.~(\ref{eq:n_bar}) and (\ref{eq:delta_q}). This is indeed the case for \ankaa qubits, as illustrated by \cref{fig:reset_calibration}(a) showing the results of two-tone spectroscopy for Qubit~76.

The two-tone spectroscopy was performed in the following way~\cite{tosi_effects_2024}. We first prepared an equal superposition of the $|g\rangle$ and $|e\rangle$ states by applying an $R_x(\pi / 2)$ rotation to the qubit initialized in the ground state. We then simultaneously drove the qubit and its readout resonator for $1.5~\mu\text{s}$ while keeping the amplitudes of the drives fixed and varying their frequencies. After we turned off the drives, we waited for 360~ns to let the resonator relax to its ground state. Finally, we measured the qubit state using a regular \ankaa readout pulse. The measurement outcomes averaged over 200 shots are shown in \cref{fig:reset_calibration}(a) by color, with red/blue corresponding to the ground/excited state. Two overlapping Lorentzian dips can be clearly seen in \cref{fig:reset_calibration}(a). Note that the Lorentzians are distorted due to a non-linear dependence of the dispersive shift on $\bar n_{g,e}$ that we have neglected~\cite{tosi_effects_2024}. We performed all our measurements using Rigetti's Quantum Cloud Services (QCS) platform, which allows for pulse-level control.

The DDROP protocol~\cite{geerlings_demonstrating_2013} is based on the same principles as the two-tone spectroscopy we just described. To reset a qubit, we apply two simultaneous drives at the frequencies corresponding to the left Lorentzian dip in \cref{fig:reset_calibration}(a). After applying the two drives for sufficiently long time (of about $10/\kappa$), the coupled qubit-resonator system is driven to the steady state $|g, \bar n_g\rangle$. Then the drives are turned off, and the system spontaneously relaxes to the target state $|g, 0\rangle$, which happens on a time scale of about $1 / \kappa$. Note that the right dip in \cref{fig:reset_calibration}(a) (shown in blue) corresponds to the steady state $|e, \bar n_e\rangle$. Therefore, by choosing an appropriate combination of drive frequencies, this protocol can also be used for bringing the qubit to the $|e\rangle$ state instead of $|g\rangle$.

We used the following procedure to coarsely calibrate unconditional resets for \ankaa. We kept the resonator drive amplitude fixed at the level used for mid-circuit measurements, which in turn was optimized to avoid leakage (as described in Methods). We then fixed the qubit drive strength at some initial value and performed two-tone spectroscopy. Based on the two-tone spectroscopy results, we chose a combination of the drive frequencies corresponding to the left Lorentzian in Fig.~\ref{fig:reset_calibration}(a) (one of the possible choices is shown by the black cross in the figure). Having fixed the amplitude of the resonator drive and the frequencies of both drives, we then performed a sweep of the qubit drive amplitude to minimize the excited-state population measured after the reset pulse. Finally, we tested the performance of the DDROP resets by varying the duration of the simultaneous drives and measuring the resulting state $|e\rangle$ population. The corresponding experimental results are shown in Fig.~\ref{fig:reset_calibration}(b) and are also fitted with an exponentially decaying function. For Qubit~76 shown in Fig.~\ref{fig:reset_calibration}(b), we extracted the steady-state population of 0.045 and the decay time constant of $0.31~\mu\text{s}$, but similar values were obtained for other qubits. Based on these results, we fixed the reset pulse duration at $1.5~\mu\text{s}$. In our experience, this time can be reduced by increasing the resonator drive strength, but this can also increase leakage (see Methods). Note that the observed steady-state population is likely limited by the measurement fidelity rather than by the performance of the DDROP resets.

\begin{figure}
\includegraphics[width=0.6\columnwidth]{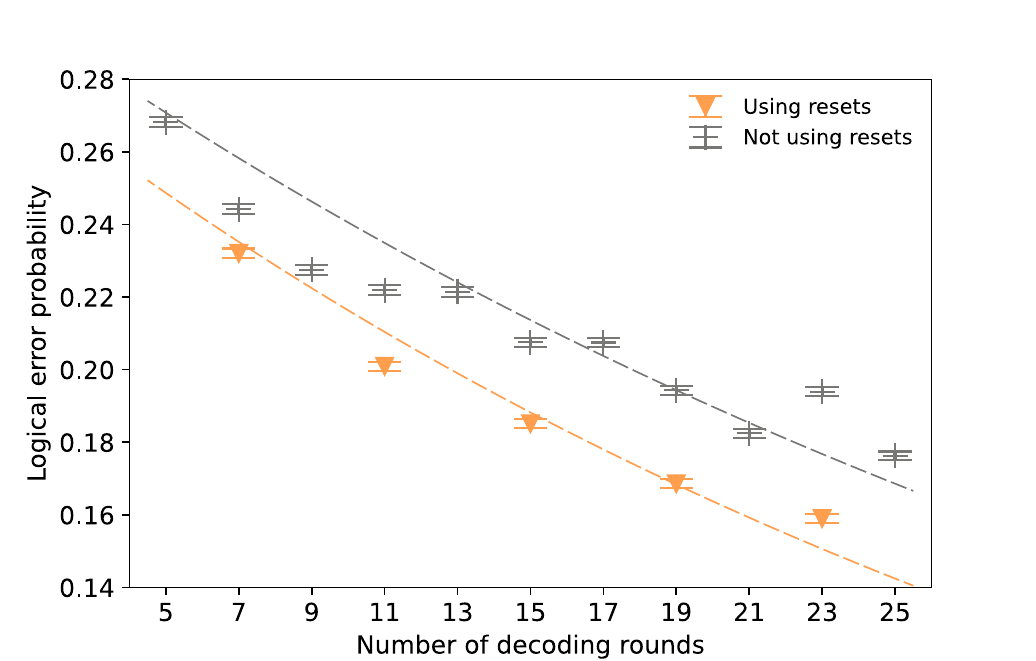}
\caption{\label{fig:resets_3}\textbf{Impact of unconditional resets on the logical error probability.} Shown is the logical error probability as a function of the number of decoding rounds for stability-8 experiments with and without using unconditional resets after each syndrome extraction round. The syndrome data are decoded with a software implementation of the MWPM decoder using soft information and the pairwise correlation method to construct a decoding graph. The error bars are computed as standard deviations of the mean values. 
}
\end{figure}

Figure~\ref{fig:resets_3} shows the logical error probability for the stability-$8$ experiment performed both with and without resetting ancillas between syndrome extraction rounds. We decoded offline with the MWPM decoder using soft information and constructing the decoding graph with the pairwise correlation method. Our results indicate that resetting the ancilla qubits improves the logical error probability. We plan to further investigate the DDROP protocol elsewhere, including the possibility to adapt it for resetting transmons in higher excited states.

\bibliography{supplement}